\documentclass[a4paper,fleqn]{cas-dc}

\usepackage[numbers]{natbib}
\usepackage{amsmath,mathtools,amssymb}

\newtheorem{theorem}{Theorem}
\newtheorem{lemma}{Lemma}

\newtheorem{proposition}{Proposition}

\newtheorem{ass}{Assumption}

\newcommand{\R}{\mathbb{R}}

\newcommand{\step}[1]{\textit{Step #1:}}

\newcommand\norm[1]{\ensuremath{\lVert#1\rVert}}

\DeclareMathOperator\diag{diag}

\def\tsc#1{\csdef{#1}{\textsc{\lowercase{#1}}\xspace}}
\tsc{WGM}
\tsc{QE}
\begin{document}
\let\WriteBookmarks\relax
\def\floatpagepagefraction{1}
\def\textpagefraction{.001}

% Short title
\shorttitle{}    

% Short author
%\shortauthors{Daniele Ravasio et~al.}

% Main title of the paper
\title [mode = title]{Development of a velocity form for a class of RNNs, with application to offset-free nonlinear MPC design}  

%Title footnote mark
% eg: \tnotemark[1]
% \tnotemark[1] 

% Title footnote 1.
% eg: \tnotetext[1]{Title footnote text}
% \tnotetext[1]{} 

% First author
%
% Options: Use if required
% eg: \author[1,3]{Author Name}[type=editor,
%       style=chinese,
%       auid=000,
%       bioid=1,
%       prefix=Sir,
%       orcid=0000-0000-0000-0000,
%       facebook=<facebook id>,
%       twitter=<twitter id>,
%       linkedin=<linkedin id>,
%       gplus=<gplus id>]

\author[1,2]{Daniele Ravasio}
%\ead{daniele.ravasio@polimi.it}
\cormark[1]
\author[1]{Bestem Abdulaziz}
%\ead{bestem.abdulaziz@mail.polimi.it}
\author[1]{Marcello Farina}
%\ead{marcello.farina@polimi.it}
\author[2]{Andrea Ballarino}
%\ead{andrea.ballarino@stiima.cnr.it}

%Footnote of the first author
%\fnmark[1]

%Email id of the first author
%\ead{}

%URL of the first author
%\ead[url]{}

%Credit authorship
%eg: \credit{Conceptualization of this study, Methodology, Software}
%\credit{}

%Address/affiliation
\affiliation[1]{organization={Dipartimento di Elettronica, Informazione e Bioingegneria (DEIB), Politecnico di Milano},
            addressline={Via Ponzio 34/5}, 
            city={Milan},
            postcode={20133}, 
            country={Italy}}
\affiliation[2]{organization={Istituto di Sistemi e Tecnologie Industriali Intelligenti per il Manifatturiero Avanzato (STIIMA) - Consiglio Nazionale delle Ricerche (CNR)},
            addressline={Via A. Corti 12}, 
            city={Milan},
            postcode={20133}, 
            country={Italy}}

% Footnote of the second author
%\fnmark[2]

% Email id of the second author
%\ead{}

% URL of the second author
%\ead[url]{}

% Credit authorship
%\credit{}

% Corresponding author text
\cortext[1]{Corresponding author}

%%%%%%%%%
\makeatletter
\def\printorcid{}
\makeatother

% Footnote text
%\fntext[1]{}

% For a title note without a number/mark
%\nonumnote{}

% Here goes the abstract
\begin{abstract}
This paper addresses the offset-free tracking problem for nonlinear systems described by a class of recurrent neural networks (RNNs). To compensate for constant disturbances and guarantee offset-free tracking in the presence of model-plant mismatches, we propose a novel reformulation of the RNN model in velocity form. A theoretically sound offset-free nonlinear model predictive control algorithm is developed. The algorithm exploits the velocity form model as the prediction model and the static controller as an auxiliary law for the definition of the terminal ingredients. Conditions based on linear matrix inequalities are derived for the design of the auxiliary control law and a nonlinear state observer.
Simulations on a pH-neutralisation process benchmark demonstrate the effectiveness of the proposed approach.
\end{abstract}

% Use if graphical abstract is present
%\begin{graphicalabstract}
%\includegraphics{}
%\end{graphicalabstract}

% Research highlights
% \begin{highlights}
% \item Proposes a physics-informed learning framework for a class of recurrent neural networks tailored to large-scale and networked systems.
% \item Enables the learning of control-oriented models that preserve the structural and stability properties of the plant.
% \item Formulates the learning problem as a convex optimisation problem with linear matrix inequality constraints to enforce desired system properties.
% \item Exploits the structural modularity of the plant to enable parallelised learning with limited communication among neighbouring subsystems.
% \end{highlights}

%\nocite{*}

% Keywords
% Each keyword is seperated by \sep
\begin{keywords}
  nonlinear model predictive control \sep recurrent neural networks \sep linear matrix inequalities \sep tracking
\end{keywords}

\maketitle

% Main text
\section{Introduction}
The use of neural networks for the design of data-driven control algorithms has attracted increasing attention in recent years~\cite{miller1995neural,tang2022data}. In the context of indirect data-based control, in particular, recurrent neural networks (RNNs) have emerged as powerful modelling tools for dynamical systems~\cite{bonassi2022recurrent}. This growing interest is due to the potential advantages of these approaches over traditional model-based ones, which rely on a model of the plant derived based on the plant physical equations. RNN models can be trained directly from plant data and subsequently employed for the design of model-based control architectures. Owing to their ability to capture long-term and nonlinear dependencies, RNNs are particularly appealing when the system under control exhibits complex nonlinear behaviour that prevents the use of standard linear model structures. \\
However, the application of such nonlinear models in practical, safety-critical settings is hindered by two key challenges: the difficulty of providing formal stability guarantees and the degradation of control performance in the presence of persistent perturbations or modelling mismatches.

Concerning the first challenge, several studies have investigated the problem of certifying stability properties of RNN models (see, e.g., \cite{buehner2006tighter,revay2020convex,bonassi2022recurrent,revay2023recurrent,RAVASIO2027106658}). Nonetheless, only a limited number of works have addressed the control design problem, i.e., the definition of tractable conditions for imparting closed-loop stability and performance guarantees to the RNN-based control system~\cite{d2023incremental,ravasio2024lmi,la2025regional,ravasio2025recurrentneuralnetworkbasedrobust}. Among these approaches,~\cite{la2025regional,ravasio2025recurrentneuralnetworkbasedrobust} propose design procedures based on linear matrix inequalities (LMIs) for the design of controllers characterised by global or regional stability and performance guarantees, and for deriving the related invariant regions where these closed-loop properties are guaranteed. The main idea in deriving these conditions is to describe the nonlinear activation functions of the RNN model through a generalised sector condition~\cite{massimetti2009linear}, i.e., a condition that allows a (possibly local) description of the nonlinearity. Notably, these approaches can also be applied to plants where global stability is not admissible. This may occur, for instance, due to the RNN’s nonlinear dynamics or to the boundedness of some state and input variables in closed-loop scenarios involving unstable dynamics.\\
These approaches, however, are developed considering shallow (i.e., single-layer) RNN models, whereas it is well known that deep (i.e., multi-layer) architectures possess superior representational capabilities for capturing complex nonlinear dynamics~\cite{lecun2015deep}. 

A second important challenge arises from the fact that the tracking performance of RNN-based control systems relies on the accuracy of the RNN model. In practice, model–plant mismatches or the presence of constant external disturbances may lead to steady-state tracking errors. 
Several offset-free strategies have been proposed in the control literature to overcome this problem and achieve perfect tracking~\cite{pannocchia2015offset}. A common approach is to augment the system with an integral action~\cite{magni2001stabilizing} and design a stabilising controller, for instance based on nonlinear model predictive control (NMPC), for the resulting augmented system. An alternative strategy for offset-free tracking is to augment the state dynamics with a disturbance model~\cite{morari2012nonlinear}. This \emph{fictitious} disturbance accounts for the mismatch between the plant and the model. A state observer is then designed for the augmented system, enabling estimation and compensation of the disturbance. \\
However, a common limitation of these strategies is that they require the computation of the state and input steady-state pair associated with the reference setpoint, which may be difficult or even impossible to determine for nonlinear models and may be uncertain. In particular, if discrepancies exist between the plant and the model, the computed steady-state target becomes inaccurate, resulting in steady-state offsets \cite{pannocchia2001velocity}.\\
A different approach, developed for linear systems, consists in reformulating the model dynamics in the so-called velocity form (see \cite{pannocchia2001velocity,wang2004tutorial,betti2013robust}), where the augmented state includes the state increments and the output tracking error, whereas the manipulated variable corresponds to the control increment. The main advantage of this formulation is that the tracking problem is recast as a regulation one, thus eliminating the need to compute the steady-state target when designing the control law. However, a major limitation is that the velocity form approach relies on the linear structure of the model, which confines its applicability to linear systems only.

In this work, we focus on a class of deep RNN models and develop a novel ad-hoc velocity form-based control scheme that enables driving the system towards a desired setpoint without requiring knowledge of the associated steady-state values. The main contributions of this paper are as follows:\\
\emph{(i)} We propose a novel approach to extend, to the considered RNN class, the model reformulation in velocity form;\\
\emph{(ii)} Extending the incremental sector condition in~\cite{ravasio2025recurrentneuralnetworkbasedrobust}, an LMI-based design procedure is developed for the considered deep RNN model. In particular, this method is used to design an offset-free nonlinear control law and a state observer that estimates the true state of the system in the presence of model-plant mismatches or constant perturbations;\\
\emph{(iii)} A theoretically sound NMPC algorithm with offset-free tracking guarantees is proposed. The algorithm uses the velocity form model as the prediction model and integrates the stabilising control law and the associated invariant set as terminal components, ensuring an enlarged region of attraction.
\smallskip\\
The rest of the paper is organised as follows. In Section~\ref{sec:problem_statement}, the considered RNN model is introduced and reformulated in velocity form. %The incremental sector condition is also presented, and the relationship between the variables of the RNN and the velocity form dynamics is analysed.
Conditions for the design of the nonlinear state-feedback control law, the offset-free NMPC law, and the nonlinear state observer are presented in Sections~\ref{sec:control_design}, \ref{sec:MPC}, and~\ref{sec:observer}, respectively. Simulation results are
provided in Section~\ref{sec:simulations}. Some final considerations are presented in Section~\ref{sec:conclusions}. The proof of the main results is provided in the Appendix.\smallskip\\
\textbf{Notation.} 
The set of real numbers is denoted by $\mathbb{R}$, while $\mathbb{R}_{\geq 0}$ denotes the set of non-negative real numbers, and $\mathbb{R}_+ \coloneq \mathbb{R}_{\geq 0} \setminus \{0\}$ denotes the set of strictly positive real numbers. Given a set $\mathcal{X}$, the notation $\mathcal{X}^n$ represents the cartesian product of $\mathcal X$ taken $n$ times. Given a vector $v \in \mathbb{R}^n$, $v^\top$ denotes its transpose, and $v_i$ its $i$-th entry. Given a matrix $M \in \mathbb{R}^{n \times n}$, its $i$-th row is denoted by $M_i$. Given $n$ matrices $M^{(1)},M^{(2)},\dots,M^{(n)}$, we denote by $\diag(M^{(1)},\dots,M^{(n)})$ the block-diagonal matrix with $M^{(1)},\dots,M^{(n)}$ on its main diagonal blocks. The matrix $I_n$ denotes the $n \times n$ identity matrix.
The set of positive definite real symmetric matrices is denoted by $\mathbb{S}_+^n \coloneq \{ M \in \mathbb{R}^{n \times n} \mid M = M^\top \succ 0 \}$, the set of diagonal positive definite matrices is defined as $\mathbb{D}_+^n \coloneq \{ M \in \mathbb{R}^{n \times n} \mid M \succ 0 \text{ and } m_{ij} = 0 \ \forall i \neq j \}$, and the set of diagonal positive semidefinite matrices is defined as $\mathbb{D}_{\geq 0}^n \coloneq \{ M \in \mathbb{R}^{n \times n} \mid M \succeq 0 \text{ and } m_{ij} = 0 \ \forall i \neq j \}$.
Given a matrix $Q \in \mathbb{S}_+^n$, we define the ellipsoidal set $\mathcal{E}(Q)$ as $\mathcal{E}(Q) = \{ v \in \mathbb{R}^n \mid v^\top Q v \leq 1 \}$.
In the following, we say that a square matrix $M \in \mathbb{R}^{n \times n}$ belongs to the set $\mathbb{B}_{\Theta}$, i.e., $M \in \mathbb{B}_{\Theta}$, if $\operatorname{rank}(I_n - \Theta M) = n$ for all $\Theta \in \mathbb{D}_+^n$ satisfying $\Theta \preceq I_n$.
The sequence $\{u(k), \dots, u(k+N)\}$ is denoted compactly as $u([k:k+N])$. Given an index set $\mathcal{I} \subseteq \{1, \dots, n\}$, let $|\mathcal{I}|$ denote the cardinality of $\mathcal{I}$. We denote by 
$M_{\mathcal{I}}\in\R^{|\mathcal{I}|\times n}$ the matrix obtained from 
$M\in\R^{n\times n}$ by removing all rows whose indices are not contained in $\mathcal{I}$. We also denote by 
$ 
[
v_i
]_{i \in \mathcal{I}}
\in \mathbb{R}^{|\mathcal{I}|}$
the column vector collecting all elements $v_i$ whose indices belong to 
$\mathcal{I}$. Let $x(k) \in \mathbb{R}^n$ denote a vector at discrete time $k$. We denote $x^+ := x(k+1)$ and $x^- := x(k-1)$. Finally, given two sets $\mathcal{A},\mathcal{B}\subseteq\mathbb{R}^n$, the Minkowski sum and Pontryagin set difference are defined as $\mathcal{A}\oplus\mathcal{B}:=\{a+b\,:\, a\in\mathcal{A},\,b\in\mathcal{B}\}$ and $\mathcal{A}\ominus\mathcal{B}:=\{x\in\mathbb{R}^n\,:\, x+\mathcal{B}\subseteq\mathcal{A}\}$, respectively.
%
%
%%%%%%%%%%%%%%%%%%%%%%%%%%%%%%%%%%%%%%%%%%%%%%%%%%%%%%%%%%%%%%%%%%%%%%%%%
\section{PROBLEM STATEMENT}\label{sec:problem_statement}
\subsection{Plant model and control goals}
In this paper, we address the problem of controlling a nonlinear plant, whose dynamics is described by the following RNN model,
\begin{equation}\label{eq:plant_dynamics}
%\scriptsize
\begin{cases}
x(k+1)=Ax(k)+Bu(k)+B_ss(k)\\
s(k)=\sigma(\tilde Ax(k)+\tilde Bu(k)+\tilde B_ss(k))\\
y(k)=Cx(k)
\end{cases}\,, 
\end{equation}
where $x\in \R^n$ denotes the state vector, $u\in\R^m$ the input vector, $y\in\R^p$ the output vector, $A\in\R^{n\times n}$, $B\in\R^{n\times m}$, $B_s\in\R^{n\times \nu}$, $\tilde A\in\R^{\nu\times n}$, $\tilde B\in\R^{\nu\times m}$, $\tilde B_s\in\R^{\nu\times \nu}$, $C\in\R^{p\times n}$, and $\sigma(\cdot)=\begin{bmatrix}\sigma_1(\cdot)&\dots&\sigma_\nu(\cdot)\end{bmatrix}^{\top}$ is a decentralized vector of scalar functions.\\
The main goal of this paper is to design a state-feedback controller to steer the plant's output $y$ to a generic constant reference signal $\bar{ y}\in\R^p$, achieving offset-free tracking. Also, we assume that the plant input and output variables are subject to constraints $$u(k)\in\mathbb{U},\qquad y(k)\in\mathbb{Y}$$ for all instants $k$, where $\mathbb{U}$ and $\mathbb{Y}$ satisfy the following assumption.
\begin{ass}\label{ass:process_constr}
    The sets $\mathbb U$ and $\mathbb Y$ are polytopes, i.e $\mathbb U=\{u\in\R^m:G_\mathrm uu\leq b_\mathrm u\}$ and $\mathbb Y=\{y\in\R^p:G_{\mathrm{y}}y\leq b_\mathrm y\}$.
\end{ass}
We also impose the following assumption on the set-point $\bar y$.
\begin{ass}\label{ass:set-point}
    The set-point $\bar y$ belongs to the output constraint set, i.e., $\bar y\in\mathbb Y$.
\end{ass}
We make the following Assumptions on model \eqref{eq:plant_dynamics}.
\begin{ass}\label{ass:sigmoid_function}
    Each component $\sigma_i:\R\rightarrow\R$, $i=1,\dots,\nu$, is %a sigmoid function, i.e., 
    a bounded, twice continuously differentiable function with positive first derivative at each point and one and only one inflection point at the origin. Also, $\sigma_i(\cdot)$ is Lipschitz continuous with unitary Lipschitz constant and
such that $\sigma_i(0)=0$, $\frac{\partial\sigma_i(v_i)}{\partial v_i}\big|_{v_i=0}=1$ and $\sigma_i(v_i)\in[-1,1]$, $\forall v_i\in\R$.
\end{ass}
Assumption~\ref{ass:sigmoid_function} is satisfied by different sigmoid-like activation functions, e.g., the hyperbolic tangent $\tanh(\cdot)$ and appropriately normalised logistic functions.
\begin{ass}\label{ass:velocity_form_ass}
For each $\Theta \in \mathbb{D}^\nu_+$ such that $\Theta \preceq I_\nu$, the following conditions hold:
\begin{enumerate}
    \item[(i)] $\operatorname{rank}(\Phi) = \nu$, where $\Phi=I_\nu - \Theta \tilde B_s$;
    \item[(ii)] $\operatorname{rank}(M) = n + p$, where
    \[
    M = \begin{bmatrix} I_n-A-B_s\Phi^{-1}\Theta\tilde A & -B-B_s\Phi^{-1}\Theta\tilde B \\ C%(A+B_s\Phi^{-1}\Theta\tilde A) 
    & 0%C(B+B_s\Phi^{-1}\Theta\tilde B)
    \end{bmatrix}.
    \]
\end{enumerate}
\end{ass}\,\\
Note that Assumption~\ref{ass:velocity_form_ass}-\emph{(i)} can be verified by properly constraining the parameter $\tilde B_\mathrm s$ during the identification phase~\cite{RAVASIO2027106658,revay2023recurrent}. For example, it is trivially satisfied by choosing $\tilde{B}_s$ as a strictly lower triangular matrix, a triangular matrix with all diagonal entries smaller than one, or a diagonally Schur stable matrix (see~\cite{farina2013block,bhaya1993discrete} for further discussion). 
Alternatively, a less restrictive condition for satisfying Assumption~\ref{ass:velocity_form_ass}-\emph{(i)} 
is stated in the following lemma.
\begin{lemma}\label{lem:Hurwitz_D_stability}
Consider a square matrix $B_s \in \mathbb{R}^{\nu \times \nu}$. If there exists a matrix $P \in \mathbb{D}_+^\nu$ such that
\begin{equation}\label{eq:digonal_hurwitz_stability}
    B_s^\top P + P B_s - 2P \prec 0\,,
\end{equation}
then, $B_s \in \mathbb{B}_\Theta$.
\hfill{}$\square$
\end{lemma}
Note that condition~\eqref{eq:digonal_hurwitz_stability} coincides with the sufficient condition used, e.g., in~\cite{revay2023recurrent}, to guarantee the well-posedness of~\eqref{eq:plant_dynamics}, i.e., the existence of a unique state trajectory for any initial condition and input sequence. \smallskip\\
%Lemma~\ref{lem:Hurwitz_D_stability} establishes that this condition also implies $B_s\in\mathbb{B}_\Theta$. %This result is exploited in the following to guarantee the well-definition of the proposed nonlinear control law in the control design.\\
%In the following, the condition in Lemma~\ref{lem:Hurwitz_D_stability} is used to guarantee the well-posedness of the nonlinear control law in the control design.
%On the other hand, Assumption~\ref{ass:velocity_form_ass}-\emph{(ii)} concerns the existence of a unique steady state associated with each value of $y$. Note, for instance, that for $\Theta=I_\nu$, matrix $M$ corresponds to the system matrix of \eqref{eq:plant_dynamics} linearised in the equilibrium $(x, u, y) = (0,0,0)$.
Assumption~\ref{ass:velocity_form_ass}-\emph{(ii)}, on the other hand, is a technical assumption with an important structural meaning. In particular, let $\bar y\in\R^p$ be a generic setpoint with $(\bar x,\bar u)$ representing the related steady-state pair. Defining $\delta x=x-\bar x$, $\delta u=u-\bar u$, and $\delta y=y-\bar y$, it is possible to show that system~\eqref{eq:plant_dynamics} linearised in $(\bar x,\bar u,\bar y)$ is %at any equilibrium point, are $A + B_s \Phi^{-1} \tilde A$ and $B + B_s \Phi^{-1} \tilde B$
\begin{equation}\label{eq:linearised_dyn}
    \begin{cases}
        \delta x(k+1){=}\bar A\delta x(k)+\bar B\delta u(k)\\
        \delta y (k)=C\delta x(k)    \end{cases}
\end{equation}
where $\bar A{=}A {+} B_s \Phi^{-1} \Theta\tilde A$ and $\bar B{=}B {+} B_s \Phi^{-1} \Theta\tilde B$, for a given $\Theta\in\mathbb D_+^\nu$, with $\Theta\preceq I_\nu$.  Consequently, matrix $M$ can be interpreted as the system matrix associated with~\eqref{eq:linearised_dyn}.  
The full-rank condition on $M$ thus ensures the existence of a unique steady-state for each  $\bar y$, which is a key requirement for the well-posedness of the velocity form representation adopted in this work. Notice that this assumption is closely related to~\cite[Assumption~1]{koelewijn2025equilibrium} and is generally satisfied when the underlying plant itself has a unique steady-state pair for each admissible output setpoint. In this case, a model identified from sufficiently informative data is expected to inherit the same property. An interesting direction for future work is to enforce this structural property directly during model training through suitable matrix inequality constraints.
 %a Schur stable symmetric matrix, a Schur stable positive matrix (i.e., a matrix with all entries strictly greater than zero), or a diagonally Schur stable matrix (see~\cite{farina2013block} and~\cite{bhaya1993discrete} for further discussion). 
 %as proposed in \cite{revay2020lipschitz}, so that there exists a diagonal matrix $W\in\mathbb{D}_+^\nu$ satisfying $2W-W\tilde B_s-\tilde B_s^\top W\succ 0$.
\smallskip\\
Finally, note that model~\eqref{eq:plant_dynamics} shares the same structure as the recurrent equilibrium network model proposed in \cite{revay2023recurrent}. However, in this work, we do not impose any open-loop stability or contractivity requirements.

\subsection{The  velocity form}
To solve the setpoint tracking problem, system \eqref{eq:plant_dynamics} is transformed in the so-called velocity form denoting $\Delta  x(k)= x(k)- x(k-1)$, $\epsilon(k)= y(k)-\bar{ y}$, $\Delta u(k)= u(k)- u(k-1)$, and $\Delta s_\mathrm c(k)=s(k)- s(k-1)$. System \eqref{eq:plant_dynamics} can thus be reformulated as
\begin{equation}\label{velocityform_dynamics}
    \begin{cases}
      \Delta x(k{+}1)= A\Delta x(k){+} B\Delta u(k){+} B_s \Delta s_\mathrm c(k)\\
      \epsilon(k{+}1){=}\epsilon(k){+} C A\Delta x(k){+} C B\Delta u(k){+} C B_s\Delta s_\mathrm c(k)
    \end{cases}
\end{equation}
Defining $\xi(k)=\begin{bmatrix}\Delta  x(k)^\top &\epsilon(k)^\top\end{bmatrix}^\top\in\R^{n_\xi}$,
\begin{equation*}
\mathcal{A}=\begin{bmatrix} A &  0_{n\times p}\\ C A &  I_p\end{bmatrix}, \
\mathcal{B}=\begin{bmatrix} B\\ C B\end{bmatrix},\ 
\mathcal{B}_s=\begin{bmatrix} B_s\\ C B_s\end{bmatrix}.
\end{equation*}
system \eqref{velocityform_dynamics} can be rewritten in compact form, i.e.,
\begin{equation}\label{eq:velocityform_dynamics_compact}
     \xi(k+1)=\mathcal{A}\xi(k)+\mathcal{B}\Delta  u(k)+\mathcal{B}_s\Delta  s_\mathrm c(k)\,.
\end{equation}
Note that, since $\xi \to 0$ implies $y \to \bar{y}$ and $\Delta x \to 0$, the tracking problem can be reformulated as the problem of regulating the state of \eqref{eq:velocityform_dynamics_compact} to the origin.\\
A key advantage of this reformulation is that it removes the need to explicitly compute the steady-state values of the plant states and inputs. This is particularly beneficial for nonlinear models, where such values can be difficult to determine. Even more so, in case of model-plant mismatches or unknown (constant) disturbances, the steady-state target computed from model \eqref{eq:plant_dynamics} may be incorrect, leading to steady-state tracking errors. In contrast, the velocity form~\eqref{eq:velocityform_dynamics_compact} guarantees offset-free behaviour, since the targets of the variables $\Delta x$ and $\epsilon$ remain zero, regardless of model-plant mismatches and constant perturbations \cite{pannocchia2001velocity}.\medskip\\
However, in order to use the form \eqref{eq:velocityform_dynamics_compact} for control design, we need to properly characterise the nonlinear relationship between the term $\Delta s_c(k)$ and states $\Delta x(k)$ and $\Delta u(k)$. As in~\cite{grimm2003establishing}, this can be done with a suitable sector condition which, although it introduces some conservativeness, is extremely beneficial for analysis and design in view of the fact that it can be formulated as a linear matrix inequality. This is done in the following Section \ref{sec:Incremental sector condition}.
\subsection{Incremental sector condition}
\label{sec:Incremental sector condition}
In order to establish the design conditions ensuring stability of the origin for \eqref{eq:velocityform_dynamics_compact}, the following lemma allows us to characterise the nonlinear relationship between the term $\Delta s_c(k)$ and states $\Delta x(k)$ and $\Delta u(k)$ in \eqref{eq:velocityform_dynamics_compact}. The characterisation introduced here extends the condition introduced in~\cite{grimm2003establishing} to a regional setting, allowing us to reduce its conservativeness if necessary.
\begin{lemma}\label{lem:sector_condition}
Define $\Delta s_i\coloneq\Delta s_i(v_i,v_i+\Delta v_i)= \sigma(v_i+\Delta v_i)-\sigma(v_i)$, where $v_i,\Delta v_i\in \R$. Under Assumption~\ref{ass:sigmoid_function}, for all $\lambda_i\in(0,1)$, $\exists\bar v_i(\lambda_i)\in\R_{+}$ such that
\begin{equation}\label{eq:sector_condition_s}
(\Delta v_i-\Delta s_i)(\Delta s_i-\lambda_i \Delta v_i)\geq0,
\end{equation}
if $v_i$ and $v_i+\Delta v_i$ are sufficiently close to $0$, i.e.,  for some $\bar{v}_i(\lambda_i)>0$, $(v_i, v_i+\Delta v_i)\in[-\bar v_i(\lambda_i),\bar v_i(\lambda_i)]^2$. Function $\bar v_i(\lambda_i):(0,1)\to(0,+\infty)$ is a continuous, strictly monotonically decreasing function such that $\bar v_i(\lambda_i) \to +\infty$ as $\lambda_i\to 0^+$ and $\bar v_i(\lambda_i)\to0$ as $\lambda_i\to 1^-$. Moreover, in case $\lambda_i=0$, condition \eqref{eq:sector_condition_s} holds for all $(v_i, v_i+\Delta v_i)\in\R^2$.%\hfill{}$\square$
\end{lemma}
The proof of Lemma~\ref{lem:sector_condition} is provided in the Appendix.\\
The incremental sector condition in Lemma~\ref{lem:sector_condition} provides a characterisation of the increment of the nonlinearity $\sigma_i(\cdot)$, for $i=1,\dots,\nu$, appearing in model~\eqref{eq:velocityform_dynamics_compact}. In particular, for $\lambda_i = 0$, condition~\eqref{eq:sector_condition_s} is globally satisfied by $\sigma_i(v_i)$, i.e., for any $ v_i, \Delta v_i \in \R^\nu$. Conversely, for $\lambda_i > 0$, condition~\eqref{eq:sector_condition_s} provides a regional characterisation of $\sigma_i( v_i)$, i.e., for any $ v_i,  v_i + \Delta v_i$ lying within the set $[-\bar v_i(\lambda_i), \bar v_i(\lambda_i)]$. Increasing $\lambda_i$, we progressively reduce the region of validity of the sector condition. In the controller design phase, the parameters $\lambda_i$, $i = 1,\dots,\nu$, will serve as additional degrees of freedom, allowing us to progressively relax the structural requirements of the design conditions, at the cost of a reduced region of stability.\smallskip\\
Note that for any given $\lambda_i\in(0,1)$, we can compute $\bar v_i(\lambda_i)$ numerically, by solving the following nonlinear optimisation problem \cite{ravasio2025recurrentneuralnetworkbasedrobust}\footnote{For odd functions, such as the hyperbolic tangent, problem~\eqref{eq:locality_constr} reduces to finding $\bar v_i\geq0$ such that
$
\left.\frac{\partial \sigma_i(v_i)}{\partial v_i}\right|_{\bar v_i}=\lambda_i
$,
which can be readily solved, e.g., by bisection.
},
\begin{equation}\label{eq:locality_constr}
\begin{aligned}
\bar v_i(\lambda_i) =& \max_{\tilde v_i} \quad \tilde v_i \\
&\text{subject to:} \\
\quad & \left.\frac{\partial \sigma_i(v_i)}{ \partial v_i}\right|_{ v_i = v_i^\star} \geq \lambda_i, \quad \forall v_i^\star \in [-\tilde v_i, \tilde v_i]
\end{aligned}
\end{equation}
For notational compactness, let $\Lambda = \diag(\lambda_1,\dots,\lambda_\nu)$, $\Delta v=\begin{bmatrix}\Delta v_1&\dots&\Delta v_\nu\end{bmatrix}^\top$, $\Delta s=\begin{bmatrix}\Delta  s_1&\dots&\Delta  s_\nu\end{bmatrix}^\top$, and define the set
$\mathcal{I}(\Lambda)\coloneq\{i\in\{1,\dots,\nu\}\ : \ \lambda_i\in(0,1)\}$.
In view of Lemma~\ref{lem:sector_condition}, for any matrix $S\in\mathbb{D}_+^\nu$, the following inequality holds
\begin{equation}\label{eq:sector_inequality}
    (\Delta  v-\Delta  s)^\top S(\Delta s-\Lambda\Delta v)\geq 0,\  \forall (v,v+\Delta v)\in\mathcal{V}(\Lambda)
\end{equation}
where the set $\mathcal{V}(\Lambda)$ is defined as
\begin{equation*}
    \mathcal{V}(\Lambda)\coloneq\{v\in\R^\nu \ : \ v_i\in[-\bar v_i(\lambda_i),\bar v_i(\lambda_i)], \ \forall i\in \mathcal{I}(\Lambda)\}.
\end{equation*}
\subsection{Relationship between velocity form states and states of the original model}
A challenging aspect of the velocity-form representation lies in translating constraints originally imposed on the system variables into equivalent constraints on $\xi$. This can be done thanks to the following proposition, which establishes a bijective correspondence between $(x,u)$ and $\xi$.
\begin{proposition}\label{prop:mapping_xi_xu}
Under Assumptions~\ref{ass:sigmoid_function} and \ref{ass:velocity_form_ass}, the following holds: 
\begin{equation}\label{eq:mapping_xi_xu}
\xi(k) = C_\xi \phi(x(k-1), u(k-1)) + C_{\bar y} \bar y\end{equation}
where
$\phi(x, u) = \begin{bmatrix} x^\top & u^\top & f_s(x, u)^\top \end{bmatrix}^\top$ and where $f_s(x, u)$ is the unique solution to the nonlinear equation
\begin{equation}\label{eq:nonlinear_function}
f_s(x, u) - \sigma(\tilde A x + \tilde B u + \tilde B_s f_s(x, u))=0,
\end{equation}
and the matrices $C_\xi$ and $C_{\bar y}$ are defined as
\[
C_\xi = 
\begin{bmatrix}
A - I_n & B & B_s \\
CA & CB & CB_s
\end{bmatrix}, 
\quad
C_{\bar y} = 
\begin{bmatrix}
0 \\
I_p
\end{bmatrix}.
\]
Moreover, given $\bar y$, the mapping $(x(k-1), u(k-1)) \mapsto \xi(k)$ in \eqref{eq:mapping_xi_xu} is bijective.%\hfill{}$\square$
\end{proposition}
The proof of Proposition~\ref{prop:mapping_xi_xu} is provided in the Appendix.
\section{Control design}\label{sec:control_design}
To address the MPC control problem, we first need to devise a suitable static state-feedback control law for \eqref{eq:velocityform_dynamics_compact} under the sector condition provided by Lemma \ref{lem:sector_condition}, which will take the role of the auxiliary control law. In this section we provide the necessary theoretical ground (see Section \ref{Static state-feedback control law}) and the corresponding design procedure (see Section \ref{Design procedure}).
%%%%%%%%%%%%%%%%%%%%%%%%%%%%%%%%%%%%%
%%%%%%%%%%%%%%%%%%%%%%%%%%%%%%%%%%%%%%%%%%%%%%%%%%%%%%%%%%%%%%
\subsection{Static state-feedback control law}
\label{Static state-feedback control law}
We introduce the control law
\begin{equation}\label{eq:control_law}
\Delta u(k)=K\xi(k)+\tilde{K}\Delta s_\mathrm c(k)\,,
\end{equation}
where $K\in\R^{m\times n_\xi}$ and $\tilde{K}\in\R^{m\times \nu}$ are the control gains to be defined in order to guarantee the asymptotic stability of the origin for \eqref{eq:velocityform_dynamics_compact}.
The closed-loop dynamics given by system \eqref{eq:velocityform_dynamics_compact} under the control law \eqref{eq:control_law} is described by
\begin{equation}\label{eq:closed_loop_velocityform}
     \xi(k+1)=\mathcal{A}_\mathrm K\xi(k)+\mathcal{B}_{s,\mathrm K}\Delta  s_\mathrm c(k)\,,
\end{equation}
where $\mathcal{A}_\mathrm K=\mathcal{A}+\mathcal{B}K$ and $\mathcal{B}_{s,\mathrm K}=\mathcal{B}_s+\mathcal{B}\tilde{K}$.\\
The following theorem provides a design condition to guarantee the asymptotic stability of the origin for system~\eqref{eq:closed_loop_velocityform}.
\begin{theorem}\label{th:asy_stability_ctr}
Consider the closed-loop dynamics \eqref{eq:closed_loop_velocityform}, and define $\tilde{\mathcal{A}}=\begin{bmatrix}\tilde A & 0\end{bmatrix}$, $\tilde{\mathcal{A}}_{\mathrm K}=\tilde{\mathcal{A}}+\tilde BK$, and  $\tilde{\mathcal{B}}_{s,\mathrm K}=\tilde B_s+\tilde B\tilde{K}\in\mathbb B_\Theta$.
Under Assumptions~\ref{ass:sigmoid_function} and \ref{ass:velocity_form_ass}, suppose that there exist matrices $P_\mathrm c \in \mathbb{S}_+^n$, $S_\mathrm c \in \mathbb{D}_{+}^\nu$, and $\Lambda_\mathrm c \in \mathbb{D}_{\geq 0}^\nu$, with $\Lambda_\mathrm c \prec I_\nu$, such that the following conditions hold
\begin{subequations}\label{eq:local_stability_conditions_ctr}
% \begin{equation}\label{eq:local_stability_LMI}
% \begin{aligned}
\begin{align}
&\begin{bmatrix}
P_\mathrm c & -\tilde{\mathcal{A}}_{\mathrm K}^\top S_\mathrm c\\
-S_\mathrm c \tilde{\mathcal{A}}_{\mathrm K} & U_{\mathrm S,\mathrm c}
\end{bmatrix} \nonumber\\
&\quad - 
\begin{bmatrix}
\mathcal{A}_\mathrm K^\top \\
\mathcal{B}_{s,\mathrm K}^\top
\end{bmatrix}
P_\mathrm c
\begin{bmatrix}
\mathcal{A}_\mathrm K & \mathcal{B}_{s,\mathrm K}
\end{bmatrix}
\succ -M_\mathrm c(\Lambda_\mathrm c)\,,\label{eq:local_stability_LMI}\\
%\end{aligned}
%\end{equation}
%\begin{equation}\label{eq:condition_implicit_function_ctr}
&    U_{\mathrm S,\mathrm c}\succ0\,,\label{eq:condition_implicit_function_ctr}
%\end{equation}
\end{align}
\end{subequations}
where $U_{\mathrm S,\mathrm c}=(I_\nu - \tilde{\mathcal{B}}_{s,\mathrm K})^\top S_\mathrm c + S_\mathrm c (I_\nu - \tilde{\mathcal{B}}_{s,\mathrm K})$, and $M_{\mathrm c}(\Lambda_\mathrm c)$ is symmetric and defined as %$U_{\mathrm c,\mathrm K}=(I_\nu - \tilde{\mathcal{B}}_{s,\mathrm K})^\top S_\mathrm c + S_\mathrm c (I_\nu - \tilde{\mathcal{B}}_{s,\mathrm K})\succ 0$ and
\[
\resizebox{\columnwidth}{!}{$
M_{\mathrm c}(\Lambda_\mathrm c){=}\!\begin{bmatrix}
        \tilde{\mathcal{A}}_{\mathrm K}^\top & \tilde{\mathcal{A}}_{\mathrm K}^\top\\\mathcal{\tilde{B}}_{s,\mathrm K}^\top & (\mathcal{\tilde{B}}_{s,\mathrm K}\!-\!I_\nu)^\top
    \end{bmatrix}\!\!
    \begin{bmatrix}
        S_\mathrm c\Lambda_\mathrm c  \!&\! 0\\
        0 \!&\! S_\mathrm c\Lambda_\mathrm c
    \end{bmatrix}\!\!
    \begin{bmatrix}
        \tilde{\mathcal{A}}_{\mathrm K} &\mathcal{\tilde{B}}_{s,\mathrm K}-I_\nu \\\tilde{\mathcal{A}}_{\mathrm K} & \mathcal{\tilde{B}}_{s,\mathrm K}
    \end{bmatrix}\!.
    $}\]
Then, the control law~\eqref{eq:control_law} is well-defined, and the origin $\xi=0$ has a nonempty region of attraction. In particular,
\begin{itemize}
     \item if $\Lambda_\mathrm c=0$, the region of attraction is $\R^n$, and the set $\mathcal{E}(P_\mathrm c/\gamma)$ is forward invariant for the closed-loop velocity form dynamics~\eqref{eq:closed_loop_velocityform} for every $\gamma\in\mathbb{R}_+$, i.e., $\xi(k)\in\mathcal{E}(P_\mathrm c/\gamma)$ implies $\xi(k+1)\in\mathcal{E}(P_\mathrm c/\gamma)$; %the origin is a globally asymptotically stable equilibrium for system \eqref{eq:closed_loop_velocityform}, i.e., for any initial condition $\xi(0) \in \mathbb{R}^{n_\xi}$, it holds that $\xi(k) \to 0$ as $k \to +\infty$; 
    \item if $\Lambda_\mathrm c\neq0$, there exists $\gamma_\mathrm c\in\R_+$ such that the set $\mathcal{E}(P_\mathrm c/\gamma_\mathrm c)$ is contained in the region of attraction, and the set $\mathcal{E}(P_\mathrm c/\gamma)$ is forward invariant for~\eqref{eq:closed_loop_velocityform} for every $\gamma\in(0,\gamma_\mathrm c]$. %for any initial condition $\xi(0)\in\mathcal{E}(P_\mathrm c/\gamma_\mathrm c)$, it holds that $\xi(k) \to 0$ as $k \to +\infty$. 
    Moreover, define
\begin{equation*}
    G_\mathrm c \coloneq 
    \begin{bmatrix}
        \tilde A_{\mathcal I(\Lambda_\mathrm c)} & \tilde B_{\mathcal I(\Lambda_\mathrm c)} & \tilde B_{s,{\mathcal I(\Lambda_\mathrm c)}} \\
        -\tilde A_{\mathcal I(\Lambda_\mathrm c)} & -\tilde B_{\mathcal I(\Lambda_\mathrm c)} & -\tilde B_{s,{\mathcal I(\Lambda_\mathrm c)}}
    \end{bmatrix},
    \bar b_\mathrm c \coloneq
    \begin{bmatrix}
        b_\mathrm c\\
        -b_\mathrm c
    \end{bmatrix}%\in\R^{2\nu},
\end{equation*}
where $b_{\mathrm c}\coloneq[\bar{v}_i(\lambda_{\mathrm c,i})]_{i\in\mathcal I(\Lambda _\mathrm c)}\in\R^{|\mathcal I(\Lambda_c)|}$. %, and $b_{\mathrm c,i} = +\infty$ otherwise, 
Then, a value of $\gamma_\mathrm c$ can be computed as
\begin{equation}\label{eq:PI_set_gamma_ctr}
\gamma_\mathrm c(\bar y)=\max_{\gamma\in\R_+}\gamma: b_{\mathrm c,i}^\star(\gamma)\leq \bar b_{\mathrm c,i}, \ \forall i=1,\dots,2|\mathcal I(\Lambda_\mathrm c)|\,,%,2\nu\,,
\end{equation}
where $b^*_{\mathrm c,i}(\gamma)$ is the solution to the following nonlinear optimisation problem
\begin{align*}
    b^*_{\mathrm c,i} &= \max_{(x,u)\in\R^{n}\times\R^m} \quad G_{\mathrm c,i} \phi(x,u) \\
    &\text{subject to:} \\
    &\ \ (C_\xi\phi(x,u) + C_{\bar y}\bar{y})^\top P_\mathrm c(C_\xi\phi(x,u) + C_{\bar y}\bar{y}) \leq \gamma \\
    %&\ \ s = \sigma(\tilde{A}x + \tilde{B}u + \tilde{B}_s s)
\end{align*}
\end{itemize}
%Moreover, for any $\gamma\in\R_+$ if $\Lambda_\mathrm c=0$, and for any $\gamma\in(0,\gamma_\mathrm c]$ if $\Lambda_\mathrm c\neq0$, the set $\mathcal{E}(P_\mathrm c/\gamma)$ is forward invariant for the closed-loop velocity form dynamics~\eqref{eq:closed_loop_velocityform}, i.e., $\xi(k)\in\mathcal{E}(P_\mathrm c/\gamma)$ implies $\xi(k+1)\in\mathcal{E}(P_\mathrm c/\gamma)$.
%\hfill{}$\square$
\end{theorem}
The proof of Theorem~\ref{th:asy_stability_ctr} is provided in the Appendix.\\
Note that, for $\Lambda_\mathrm c=0$ it holds that $M_{\mathrm c}(\Lambda_\mathrm c)=0$ and \eqref{eq:local_stability_LMI} guarantees global asymptotic stability of the origin of~\eqref{eq:velocityform_dynamics_compact}. On the other hand, when $\Lambda_\mathrm c \neq 0$, condition~\eqref{eq:local_stability_LMI} is relaxed, but it guarantees just local asymptotic stability of the origin.
Note that, once $\Delta u(k)$ is computed based on \eqref{eq:control_law}, the effective input injected into \eqref{eq:plant_dynamics} is \[u(k)=u(k-1)+\Delta u(k)\,.\]
\subsection{Design procedure}
\label{Design procedure}
To guarantee stability, the 
control gains $(K, \tilde{K})$ must be selected in accordance with Theorem~\ref{th:asy_stability_ctr}. %Theorems~\eqref{th:asy_stability_obs} and~\eqref{th:asy_stability_ctr}, respectively.
Nevertheless, since the associated conditions~\eqref{eq:local_stability_conditions_ctr} are not LMIs, finding a solution can be challenging. To overcome this difficulty, we propose the following LMI-based iterative heuristic procedure to determine the design parameters.\smallskip\\
%\begin{description}
\step{1} Simulate system \eqref{eq:plant_dynamics} using the dataset input sequence $\mathcal{U}_\mathrm d = \{u_\mathrm d(0), \dots, u_\mathrm d(N_\mathrm d)\}$, resulting in the  trajectories $\{x_\mathrm d(k)\}_{k=0}^{N_\mathrm d}$ and $\{s_\mathrm d(k)\}_{k=0}^{N_\mathrm d}$. Using these sequences, compute the empirical regional bound $\bar v_{\mathrm d} = [\bar v_{\mathrm d,1}, \ \dots, \ \bar v_{\mathrm d,\nu}]^\top$, where, for each $i = 1, \dots, \nu$,
\begin{align*}
    \bar v_{\mathrm d,i} = \max_{k = 0, \dots, n_d} \left( \tilde A_i x_\mathrm d(k) + \tilde B_i u_\mathrm d(k) + \tilde B_{s,i} s_\mathrm d(k) \right).
\end{align*}
\step{2} Compute $\Lambda_\mathrm d = \diag(\lambda_{\mathrm d,1}, \dots, \lambda_{\mathrm d,\nu})$, where each $\lambda_{\mathrm d,i}$ is defined as
\[
\lambda_{\mathrm d,i} = \min_{\lambda_i \in (0,1)} \lambda_i : \bar{v}(\lambda_i) \leq \bar{v}_{\mathrm d,i}\,,
\]
and set $\Lambda_\mathrm c=\Lambda_\mathrm d$.\\
\step{3} Solve the following LMI problem
\begin{subequations}
\begin{align}
&\max_{\substack{
\beta \in \mathbb{R}, \
Q_\mathrm c \in \mathbb{S}_{+}^{n_\xi}, Z \in \mathbb{R}^{m \times n_\xi}, \\
\tilde{Z} \in \mathbb{R}^{m \times \nu}, \
U_\mathrm c \in \mathbb{D}_{+}^\nu
}} \quad \beta \nonumber \\
&\text{subject to} \nonumber\\
&
\scriptsize
\begin{bmatrix}
Q_\mathrm c & -Q_\mathrm c \tilde{\mathcal{A}}^\top {-} Z^\top \tilde{B}^\top & Q_\mathrm c \mathcal{A}^\top {+} Z^\top \mathcal{B}^\top \\
{-}\tilde{\mathcal{A}} Q_\mathrm c {-} \tilde{B} Z^\top & U_{\mathrm Z,\mathrm c} & U_\mathrm c \mathcal{B}_s^\top {+} \tilde{Z}^\top \mathcal{B}^\top \\
\mathcal{A} Q_\mathrm c {+} \mathcal{B} Z & \mathcal{B}_s U_\mathrm c {+} \mathcal{B} \tilde{Z} & Q_\mathrm c
\end{bmatrix} \!\!\succeq\! \beta I
\label{eq:LMI_global_ctr}\\
&
%\scriptsize
\ U_{\mathrm Z,\mathrm c}\succ 0\label{eq:LMI_implicit_function_ctr}
\end{align}
\end{subequations}
where 
\(
U_{\mathrm Z,\mathrm c} = U_\mathrm c(I_\nu-\tilde{B}_s)^\top - \tilde{Z} \tilde{B}^\top + (I_\nu-\tilde{B}_s) U_\mathrm c - \tilde{B} \tilde{Z}\,,\) 
and set $K=ZQ_\mathrm c^{-1}$ and $\tilde{K}=\tilde{Z}U_\mathrm c^{-1}$.\\
\step{4} Solve the LMI problem
    \begin{equation}\label{opt:local_stability_LMI_opt}
    \begin{aligned}
        &\max_{P_\mathrm c\in \mathbb{S}_+^{n_\xi},S_\mathrm c\in \mathbb{D}_+^\nu,\alpha\in\R_+}\quad \alpha\\
        &\text{subject to}\\
        & \quad \eqref{eq:local_stability_LMI}\\
        &\quad f_{\text{LMI}}(P_\mathrm c)\succeq\alpha I
    \end{aligned}
    \end{equation}
    where the function $f_{LMI}(P_\mathrm c)$ can be selected by the designer to appropriately shape the invariant set, e.g. to maximise its volume (see \cite{boyd1994linear} for a more detailed discussion on possible function choices).\\
\step{5} If problem~\eqref{opt:local_stability_LMI_opt} is feasible, compute $\gamma_\mathrm c$ as defined in~\eqref{eq:PI_set_gamma_ctr}; otherwise, set $\Lambda_\mathrm c \gets \Lambda_\mathrm c + \epsilon_\mathrm c I_\nu$, where $\epsilon_\mathrm c \in \mathbb{R}_+$ is a small positive scalar, and return to Step~4.\smallskip\\
%\end{description}
Steps~1 and 2 initialise $\Lambda_\mathrm c$ from data in such a way that, if the outlined procedure is feasible for $\Lambda_\mathrm c=\Lambda_\mathrm d$, the resulting feasibility region of the control scheme is sufficiently large to enable the tracking of all setpoints in the dataset.
Also, from a practical perspective, limiting $\Lambda_\mathrm c$ such that $\Lambda_\mathrm c\succeq\Lambda_\mathrm d$ ensures that the system reliably operates within the range of data used for model identification. Note that, alternatively, we can initialise $\Lambda_\mathrm{c}=0$ to search for a global solution.\\ 
Steps~3–5 compute the control gains $(K,\tilde K)$ and the associated invariant set $\mathcal{E}(P_\mathrm c/\gamma_\mathrm c)$ in accordance with Theorem~\ref{th:asy_stability_ctr}. 
In particular, note that condition \eqref{eq:LMI_global_ctr} in Step~3 for $\beta = 0$ can be derived by applying the Schur complement to \eqref{eq:local_stability_LMI}, under the assumption $M_{\mathrm c}(\Lambda_\mathrm c) = 0$, and by substituting $Q_\mathrm c=P_\mathrm c^{-1}$, $U_\mathrm c=S_\mathrm c^{-1}$, $Z = KQ_\mathrm c$ and $\tilde{Z} = \tilde{K}U_\mathrm c$. By permitting $\beta$ to assume values smaller than zero, this condition is relaxed, thereby enabling the design of a control system with regional stability properties. Note that we solve this condition maximising $\beta$ so as to obtain a feasible solution characterised by the largest region of attraction. %\textcolor{green}{La condizione globale si ottiene ponendo $\beta = 0$ (poiché per $\Lambda_\mathrm{c} = 0$ si ha $M(0) = 0$). Tuttavia, non è sempre vero che massimizzando $\beta$ si massimizzi la regione in cui si garantisce la stabilità. Ad esempio, avevamo visto che $M_\mathrm{c}(\Lambda_\mathrm{c})$ può essere definita negativa per alcuni valori di $\Lambda_\mathrm{c}$.} 
Moreover, note that condition~\eqref{eq:LMI_implicit_function_ctr} is equivalent to~\eqref{eq:condition_implicit_function_ctr}, and therefore it guarantees that $\tilde{\mathcal{B}}_{s,\mathrm K} \in \mathbb{B}_\Theta$, i.e., that the control law~\eqref{eq:control_law} is well-defined.
\\
In Step~4, the gains $K$ and $\tilde{K}$ are fixed, so that we can solve \eqref{eq:local_stability_LMI} as an LMI problem.\\
Finally, in Step~5, if \eqref{opt:local_stability_LMI_opt} is feasible, we compute $\gamma_\mathrm c$ such that the invariant set $\mathcal{E}(P_\mathrm c / \gamma_\mathrm c)$ satisfies the locality constraints. Otherwise, the region over which stability is to be enforced is progressively reduced by updating $\Lambda_\mathrm c$, until~\eqref{eq:local_stability_LMI} is satisfied.
%
%
%%%%%%%%%%%%%%%%%%%%%%%%%%%%%%%%%%%%%%%%%%%%%%%%%%%%%%%%%%%%%%%%%%%%%%%%%%%%%
%
\section{MPC control design}\label{sec:MPC}
In this section we show how to use model~\eqref{eq:velocityform_dynamics_compact} and the control law~\eqref{eq:control_law} as the prediction model and auxiliary law in the design of an offset-free NMPC algorithm.
\subsection{Velocity form NMPC design}
To address the offset-free tracking MPC problem, the following finite-horizon optimal control problem (FHOCP) is formulated
\begin{subequations}\label{opt:NMPC}
\begin{align}
&\min_{\Delta u([k:k+N-1])} J\left(\xi([k:k+N]),\Delta u([k:k+N-1])\right) \notag\\
&\text{subject to:} \notag\\
&\xi(k)=\begin{bmatrix}x(k)-x(k-1)\\y(k)-\bar y\end{bmatrix} \label{opt:NMPC_init}\\
&\forall\tau=0,\dots,\ N-1:\notag\\
&\quad \begin{aligned} \xi(k+\tau+1)&=\mathcal{A}\xi(k+\tau)+\mathcal{B}\Delta  u(k+\tau)\\&+\mathcal{B}_s\Delta  s_\mathrm c(k+\tau)
\end{aligned} \label{opt:NMPC_system}\\
&\quad u(k-1)+\sum_{j=0}^\tau\Delta u(k+j) \in \mathbb U\label{opt:NMPC_input}\\
&\quad y(k)+[C \ 0]\sum_{j=1}^\tau\xi(k+j)\in \mathbb Y\label{opt:NMPC_outputcstr}\\
&\hat{\xi}(k+N) \in \mathbb E_\mathrm f\label{opt:NMPC_terminal}
\end{align}
\end{subequations}
Constraint \eqref{opt:NMPC_system} embeds the dynamics of the velocity form predictive model, which is initialised by constraint \eqref{opt:NMPC_init} using the most recent state and output measurements. Input and output constraints are enforced through~\eqref{opt:NMPC_input} and~\eqref{opt:NMPC_outputcstr}, respectively.  Moreover, the terminal constraint \eqref{opt:NMPC_terminal} ensures that the state $\xi$ at the end of the prediction horizon lies within the terminal set $\mathbb E_\mathrm f$.\\
Finally, the cost function is defined as
\[J=\sum_{\tau=0}^{N-1}\left(\norm{\xi(k+\tau)}_Q^2+\norm{\Delta u(k+\tau)}_R^2\right)+V_\mathrm f(\xi(k+N)),\]
where $Q\in\R^{n_\xi\times n_\xi}$ and $R\in\R^{m\times m}$ are positive definite matrices, and $V_\mathrm f$ is the terminal cost that will be specified below.
\\
The solution to the FHOCP \eqref{opt:NMPC} at time $k$ is denoted $\Delta u([k:k+N-1]|k)$.\\
As common in receding-horizon schemes, only the first input $\Delta u(k|k)$ is used to compute the control action $u(k) = u(k-1) + \Delta u(k|k)$, which is then applied to the plant. This procedure is repeated at each time step.
\subsection{Terminal ingredients}
To ensure stability of the NMPC scheme, we use \eqref{eq:control_law} as an auxiliary control law for the definition of the terminal ingredients. Exploiting its invariance properties, the terminal set is defined as
\[
\mathbb{E}_\mathrm{f} \coloneq \mathcal{E}(P_\mathrm{f}/\gamma_\mathrm{f}),
\]
where \(P_\mathrm{f} \in \mathbb{S}_+^{n_\xi}\) and \(\gamma_\mathrm{f} \in \mathbb{R}_+\) are determined according to Theorem~\ref{th:asy_stability_ctr}, while additionally ensuring that, whenever the state \(\xi(k)\) of the velocity form system lies within \(\mathbb{E}_\mathrm{f}\), the process constraints \((y,u) \in \mathbb{Y} \times \mathbb{U}\) are satisfied.\\
The terminal cost is defined as
\[
V_\mathrm{f}(\xi(k+N)) \coloneq \|\xi(k+N)\|_{P_\mathrm{f}}^2,
\]
ensuring that, under \eqref{eq:control_law}, the condition
\(
V_\mathrm{f}(\xi(k+1)) - V_\mathrm{f}(\xi(k)) \le \|\xi(k)\|_Q + \|\Delta u(k)\|_R
\)
is satisfied.\\
To satisfy these requirements, the design parameters $K$, $\tilde K$, $P_\mathrm f$, and $\gamma_\mathrm f$ are determined following the procedure outlined in Section~\ref{sec:control_design}, with steps 4 and 5 replaced by the following:\smallskip\\
\step{4-b} Set $\Lambda_\mathrm f=\Lambda_\mathrm d$ and solve the LMI problem
\begin{subequations}\label{opt:MPC_stability_LMI}
\begin{align}
    &\max_{P_\mathrm f \in \mathbb{S}_+^{n_\xi},\, S_\mathrm f \in \mathbb{D}_+^\nu,\, \alpha \in \mathbb{R}_+} \quad \alpha \nonumber\\
    &\text{subject to}\nonumber\\
    &
    \begin{aligned}
        & \begin{bmatrix}
            P_\mathrm f-Q-K^\top RK & -\tilde{\mathcal{A}}_\mathrm K^\top S_\mathrm f-K^\top R\tilde K \\
            -S_\mathrm f \tilde{\mathcal{A}}_\mathrm K-\tilde K^\top RK & U_{\mathrm S,\mathrm f}
        \end{bmatrix} \\
        &\quad - 
        \begin{bmatrix}
            \mathcal{A}_\mathrm K^\top \\
            \mathcal{B}_{s,\mathrm K}^\top
        \end{bmatrix}
        P_\mathrm f
        \begin{bmatrix}
            \mathcal{A}_\mathrm K & \mathcal{B}_{s,\mathrm K}
        \end{bmatrix}
        \succeq -M_{\mathrm c}(\Lambda_\mathrm f)
    \end{aligned} \label{eq:terminal_set_LMI} \\
    & f_{\text{LMI}}(P_\mathrm f) \succeq \alpha I 
\end{align}
\end{subequations}
where $U_{\mathrm S,\mathrm f}\coloneq(I_\nu - \tilde{\mathcal{B}}_{s,\mathrm K})^\top S_\mathrm f + S_\mathrm f (I_\nu - \tilde{\mathcal{B}}_{s,\mathrm K})- \tilde K^\top R\tilde K$. If problem~\eqref{opt:MPC_stability_LMI} is feasible, proceed to Step~5-b. Otherwise, update $\Lambda_\mathrm f \gets \Lambda_\mathrm f + \epsilon_\mathrm f I_\nu$, where $\epsilon_\mathrm f \in \mathbb{R}_+$ is a small positive scalar, and resolve \eqref{opt:MPC_stability_LMI}. This step is repeated until a feasible solution is found.\\
\step{5-b} Define
\[
G_{\mathrm{u_y}} = 
\begin{bmatrix}
0 & G_\mathrm u & 0\\G_{\mathrm{y}}C& 0& 0
\end{bmatrix}, \,
G_\mathrm f = 
\begin{bmatrix}
G_{\mathrm c}\\G_{\mathrm u_\mathrm y}
\end{bmatrix},\,
\bar b_{\mathrm f} = 
\begin{bmatrix}
\bar b_\mathrm c\\b_\mathrm u\\b_\mathrm y
\end{bmatrix}\in \R^{n_\mathrm c}\,,
\]
and compute $\gamma_\mathrm f$ by solving 
\begin{equation}\label{eq:PI_set_gamma_mpc}
\gamma_\mathrm f(\bar y)=\max_{\gamma\in\R_+}\gamma\ : b_{\mathrm f,i}^\star(\gamma)\leq \bar b_{\mathrm f,i}, \ \forall i=1,\dots,n_\mathrm c\,,
\end{equation}
where $b_{\mathrm f,i}^\star(\gamma)$ denotes the solution to the following nonlinear optimisation problem
\begin{align*}
    b_{\mathrm f,i}^\star = &\max_{(x,u)\in\R^{n}\times\R^m}\quad G_{\mathrm f,i} \phi(x,u) \\
    &\text{subject to:} \\
    &\ \ (M\phi(x,u) + L\bar{y})^\top P_\mathrm f(M\phi(x,u) + L\bar{y}) \leq \gamma \\
    %&\ \ s = \sigma(\tilde{A}x + \tilde{B}u + \tilde{B}_s s)
\end{align*}
%Note that Step~5-b ensures that whenever the state of the velocity form system lies within the invariant set, both the locality constraints and the process constraints $(y,u) \in \mathbb{Y} \times \mathbb{U}$ are satisfied.\smallskip\\
%It is worth noting that, when the auxiliary control law~\eqref{eq:control_law} provides only local stability, the associated locality constraints on $x$ and $u$ enter the NMPC problem~\eqref{opt:NMPC} exclusively through the terminal set. Hence, they do not need to be enforced along the entire prediction horizon to guarantee stability. This, in turn, allows for an enlargement of the controller’s feasibility region.
%
\subsection{Main result}
The main result, stating the properties of the MPC-based control scheme, can now be proved.
\begin{theorem}\label{th:convergence_MPC}
    Suppose that Assumptions~\ref{ass:process_constr}-\ref{ass:velocity_form_ass} are verified.
    If a solution to the FHOCP~\eqref{opt:NMPC} exists at time $k=0$, the FHOCP~\eqref{opt:NMPC} admits a solution at all $k\geq 0$, and the resulting NMPC controller asymptotically steers the system output $y$ to the desired set-point $\bar y$, while respecting the constraints $(u(k),y(k))\in\mathbb U\times\mathbb Y$ for all $k\geq0$.%\hfill{}$\square$
\end{theorem}
The proof of Theorem~\ref{th:convergence_MPC} is provided in the Appendix.
%
%
%%%%%%%%%%%%%%%%%%%%%%%%%%%%%%%%%%%%%%%%%%%%%%%%%%%%%%%%%%%%%%%%%%%%%%%%%%%%%
%
%
\section{State observer design}\label{sec:observer}
In Sections~\ref{sec:control_design} and \ref{sec:MPC} we addressed the offset-free tracking problem implicitly assuming (e.g., as in \eqref{opt:NMPC_init}) that the state of the system is available. However, the state-feedback assumption is often unrealistic in practical scenarios, especially when working with data-driven models where the system state typically does not correspond to directly measurable physical quantities. To address this potential limitation, in this section, we propose a state observer that estimates the true state of system~\eqref{eq:plant_dynamics}, accounting for perturbations and disturbances.
The observer is based on an augmented formulation of the system dynamics~\eqref{eq:plant_dynamics}, incorporating an output disturbance model \cite{pannocchia2003disturbance}, i.e.,
\begin{equation}\label{eq:disturbance_model}
\begin{cases}
        x(k+1)=Ax(k)+Bu(k)+B_ss(k)\\
        s(k)=\sigma(\tilde Ax(k)+\tilde Bu(k)+\tilde B_ss(k))\\
        d(k+1)=d(k)\\
        y(k)=Cx(k)+d(k)
\end{cases}
\end{equation}
where the “fictitious” disturbance $d\in\R^p$ is introduced, in particular, to take into account the differences between the plant and the model.\\
Define the enlarged state $\eta(k)=[x(k)^\top,\ d(k)^\top]^\top\in \R^{n_{\eta}}$ and matrices 
\[
\begin{matrix}
     &  A_\mathrm e=\begin{bmatrix}
         A & 0\\0 & I_p
     \end{bmatrix},
     &  B_\mathrm e=\begin{bmatrix}
         B\\0
     \end{bmatrix},
     &  B_{s,\mathrm e}=\begin{bmatrix}
         B_s \\ 0
     \end{bmatrix},\\
     & \tilde{A}_\mathrm e=\begin{bmatrix}
         \tilde A&0
     \end{bmatrix}, 
     & C_\mathrm e=\begin{bmatrix}
         \tilde C&I_p
     \end{bmatrix}
\end{matrix}\,.
\]
System \eqref{eq:disturbance_model} can be rewritten in compact form as 
\begin{equation}\label{eq:disturbance_model_compact}
\begin{cases}
        \eta(k+1)=A_\mathrm e\eta(k)+B_\mathrm eu(k)+B_{s,\mathrm e}s(k)\\
        s(k)=\sigma(\tilde A_\mathrm e\eta+\tilde Bu+\tilde B_ss(k))\\
        y(k)=C_\mathrm e\eta(k)
\end{cases}
\end{equation}
The proposed state observer for the augmented system reads as follows
\begin{equation}\label{eq:state_observer}
\begin{cases}
        \hat\eta(k{+}1){=}A_\mathrm e\hat\eta(k){+}B_\mathrm eu(k){+}B_{s,\mathrm e}\hat s(k){+}Le_y(k)\\
        \hat s(k){=}\sigma(\tilde A_\mathrm e\hat\eta(k){+}\tilde Bu(k){+}\tilde B_{s}\hat s(k){+}\tilde Le_y(k))
\end{cases}
\end{equation}
where $\hat \eta(k)=[\hat x(k)^\top,\ \hat d(k)^\top]^\top\in\R^{n_\mathrm e}$ is the observer state, $e_y(k)=y(k)-C_\mathrm e\hat\eta(k)$ is the innovation, and  $L\in\R^{n_\mathrm e\times q}$ and $\tilde L\in\R^{\nu\times q}$ are the observer gains, to be defined according to the following theorem.
\begin{theorem}\label{th:asy_stability_obs}
Consider the observer dynamics \eqref{eq:state_observer}, and define the estimation error as $e(k)\coloneq \eta(k)-\hat \eta(k)$. Moreover, define $A_{\mathrm e,\mathrm L}\coloneq A_{\mathrm e}-C_\mathrm eL$ and $\tilde A_{\mathrm e,\mathrm L}\coloneq\tilde A_{\mathrm e}-C_\mathrm e\tilde L$.
Under Assumption~\ref{ass:sigmoid_function}, suppose that there exist matrices $P_\mathrm o\in\mathbb S_+^{n_\mathrm e}$, $S_\mathrm o\in\mathbb{D}_{+}^\nu$, and %$ \Lambda_\mathrm o\in\mathbb{D}_{\geq0}^\nu$
$\Lambda_\mathrm o=\diag(\lambda_{\mathrm o,1},\dots,\lambda_{\mathrm o,\nu})\in\mathbb{D}_{\geq0}^\nu$, with $ \Lambda_\mathrm o\prec I_\nu$, such that the following conditions hold
\begin{subequations}\label{eq:local_stability_LMI_obs}
\begin{align}
&
\begin{bmatrix}
    P_\mathrm{o} 
    &
    -\tilde A_{\mathrm{e,L}}^\top S_\mathrm{o}
    \\
    -S_\mathrm{o}\tilde A_{\mathrm{e,L}}
    &
    (I_\nu-\tilde B_s)^\top S_\mathrm{o}
    + S_\mathrm{o}(I_\nu-\tilde B_s)
\end{bmatrix}
\nonumber\\[-1mm]
&\qquad
-
\begin{bmatrix}
    A_{\mathrm{e,L}}^\top\\
    B_{s,\mathrm{e}}^\top
\end{bmatrix}
P_\mathrm{o}
\begin{bmatrix}
    A_{\mathrm{e,L}} & B_{s,\mathrm{e}}
\end{bmatrix}
\succ -M_\mathrm{o}(\Lambda_\mathrm{o}),
\label{eq:local_stability_LMI_obs_1}
\\[2mm]
&
\begin{bmatrix}
    P_\mathrm{o}/\gamma_\mathrm{o}
    & C^\top\tilde L_i^\top
    \\
    \tilde L_iC
    & \bar v(\lambda_{\mathrm{o},i})^2
\end{bmatrix}
\succeq 0,
\qquad
\forall i\in\mathcal I(\Lambda_\mathrm{o}),
\label{eq:local_stability_LMI_obs_2}
\\[2mm]
&
\begin{bmatrix}
    P_\mathrm{o}/\gamma_\mathrm{o}
    & \tilde A_{\mathrm e,i}^\top
    \\
    \tilde A_{\mathrm e,i}
    & \bar v(\lambda_{\mathrm{o},i})^2
\end{bmatrix}
\succeq 0,
\qquad
\forall i\in\mathcal I(\Lambda_\mathrm{o}).
\label{eq:local_stability_LMI_obs_3}
\end{align}
\end{subequations}
where $M_{\mathrm o}(\Lambda_\mathrm o)$ is symmetric and defined as
\[
\resizebox{\columnwidth}{!}{$
M_{\mathrm o}(\Lambda_\mathrm o)=\begin{bmatrix}
        \tilde A_{\mathrm e,\mathrm L}^\top & \tilde A_{\mathrm e,\mathrm L}^\top\\\tilde B_{s}^\top & (\tilde B_{s}\!-\!I_\nu)^\top
    \end{bmatrix}\!\!
    \begin{bmatrix}
        S_\mathrm o\Lambda_\mathrm o & 0\\
        0 & S_\mathrm o\Lambda_\mathrm o
    \end{bmatrix}\!\!
    \begin{bmatrix}
        \tilde A_{\mathrm e,\mathrm L} & \tilde B_{s}I_\nu\\\tilde A_{\mathrm e,\mathrm L}& \tilde B_{s}
    \end{bmatrix}.
    $}\]
%Then, the observation error $e(k)=\eta(k)-\hat \eta(k)\to0$ as $k\to\infty$,
Then, the origin of the estimation error dynamics has a nonempty region of attraction. In particular,
\begin{itemize}
     %\item for any $e(0)\in\R^{n_\mathrm e}$ if $\Lambda_\mathrm o=0$;
     \item if $\Lambda_\mathrm{o}=0$, the region of attraction is $\R^{n_\mathrm{e}}$, and, for every $\gamma\in\R_+$, the set $\mathcal{E}(P_\mathrm{o}/\gamma)$ is forward invariant for the estimation error dynamics;
\item if $\Lambda_\mathrm o\neq 0$ and 
\begin{subequations}\label{eq:obs_regional_constr}
\begin{align}
&\tilde A\hat x {+} \tilde B u {+} \tilde B_s \hat s
\in \mathcal{V}(\Lambda_\mathrm{o})
\ominus \tilde L C_\mathrm{e}\mathcal{E}(P_\mathrm{o}/\gamma_\mathrm{o}),
\label{eq:obs_regional_constr_1}\\
&\tilde A\hat x {+} \tilde B u {+} \tilde B_s s
\in \mathcal{V}(\Lambda_\mathrm{o})
\ominus \tilde A_\mathrm e\mathcal{E}(P_\mathrm{o}/\gamma_\mathrm{o}),
\label{eq:obs_regional_constr_2}
\end{align}
\end{subequations}
where %$\Delta\hat s\coloneq s-\hat s$, and 
the sets 
$\mathcal{V}(\Lambda_\mathrm o) \ominus \tilde LC_\mathrm e\mathcal{E}(P_\mathrm{o}/\gamma_\mathrm{o})$ and $\mathcal{V}(\Lambda_\mathrm o) \ominus \tilde A_\mathrm e\mathcal{E}(P_\mathrm{o}/\gamma_\mathrm{o})$ are non-empty, 
then $\mathcal E(P_\mathrm o/\gamma_\mathrm o)$ is contained in the region of attraction of the origin, and, for every $\gamma\in(0,\gamma_\mathrm o]$, the set $\mathcal E(P_\mathrm o/\gamma)$ is forward invariant for the estimation error dynamics.
\end{itemize}
%Moreover, for any $\gamma\in\R_+$ if $\Lambda_\mathrm o=0$, and for any $\gamma\in(0,\gamma_\mathrm o]$ if $\Lambda_\mathrm o\neq0$, the set $\mathcal{E}(P_\mathrm o/\gamma)$ is forward invariant for the state estimation error dynamics, i.e., $e(k)\in\mathcal{E}(P_\mathrm o/\gamma)$ implies $e(k+1)\in\mathcal{E}(P_\mathrm o/\gamma)$.
%\hfill{}$\square$
\end{theorem}
The proof of Theorem~\ref{th:asy_stability_obs} is provided in the Appendix.\\
Note that if the observer~\eqref{eq:state_observer} is designed with regional stability properties, it is necessary to ensure that constraint~\eqref{eq:obs_regional_constr} is satisfied. In this regard, two cases can be distinguished.\\
\emph{(i)} If the observer can be initialised such that $e(0)$ is sufficiently small, i.e., $\eta(k)\approx\hat\eta(k)$, for instance by running the observer for a sufficiently long time before activating the controller, the effect of the estimation error can be neglected. In this case, constraints \eqref{eq:obs_regional_constr_1}-\eqref{eq:obs_regional_constr_2} reduce to $\tilde Ax+\tilde Bu+\tilde B_ss\in\mathcal V(\Lambda_\mathrm o)$. %When the control scheme based on the static control law is considered, this condition can be enforced by modifying~\eqref{eq:PI_set_gamma_ctr} to additionally ensure that when $\xi\in\mathcal E(P_\mathrm c/\gamma_\mathrm c)$, $|\tilde A_ix+\tilde B_iu+\tilde B_{s,i}s|\leq\bar v_i(\lambda_{\mathrm o,i})$ for all $i\in\mathcal I(\Lambda_\mathrm o)$. Conversely, if the NMPC-based control scheme is adopted, the same 
This condition can be explicitly enforced within the NMPC problem~\eqref{opt:NMPC}. \\
\emph{(ii)} If, instead, the initial estimation error is significant, its effect may potentially compromise the stability properties of the overall control system. Nonetheless, by Theorem~\ref{th:asy_stability_obs}, if $e(0) \in \mathcal{E}(P_\mathrm o / \gamma)$, then $e(k) \in \mathcal{E}(P_\mathrm o / \gamma)$ for all $k\in\mathbb Z_{\geq0}$. Hence, the estimation error can be regarded as a bounded disturbance, and a robust control scheme for the estimated state dynamics, expressed in velocity form, can be designed to account for its effect, following, e.g., the approach in~\cite{ravasio2025recurrentneuralnetworkbasedrobust}.
Similarly, constraints~\eqref{eq:obs_regional_constr_1}-\eqref{eq:obs_regional_constr_2} can be incorporated into the NMPC problem~\eqref{opt:NMPC} as additional constraints. They can be respectively enforced by imposing that %for %$\xi\in\mathcal E(P_\mathrm c/\gamma_\mathrm c)$,
$|\tilde A_i\hat x+\tilde B_iu+\tilde B_{s,i}\hat s|\leq\bar v_i(\lambda_{\mathrm o,i})-\max_{e\in\mathcal E(P_\mathrm o/\gamma)} |\tilde L_iC_\mathrm ee|$ and $|\tilde A_i\hat x+\tilde B_iu+\tilde B_{s,i}\hat s+w_{s,i}(\hat x,u)|\leq\bar v_i(\lambda_{\mathrm o,i})-\max_{e\in\mathcal E(P_\mathrm o/\gamma )} |\tilde A_{\mathrm e,i}e|$ for all $i\in\mathcal I(\Lambda_\mathrm o)$, where $w_{s,i}(\hat x,u)$ accounts for the effect of the estimation error on the nonlinear term and is defined for all $i\in\mathcal I(\Lambda_\mathrm o)$ as
\begin{align*}
&w_{s,i}(\hat x, u)\coloneq \max_{e\in\mathcal E(P_\mathrm o/\gamma_\mathrm o)}|\tilde B_{s,i}\Delta s|\\
&\text{subject to:  }
\Delta s=\sigma(\tilde A\hat x+\tilde Bu+\tilde B_s\hat s+\tilde B_s\Delta s+\tilde A_\mathrm ee)
\end{align*}
\section{Case study}\label{sec:simulations}
\begin{figure}%[htbp]
     \centering
    \includegraphics[width=0.9\columnwidth]{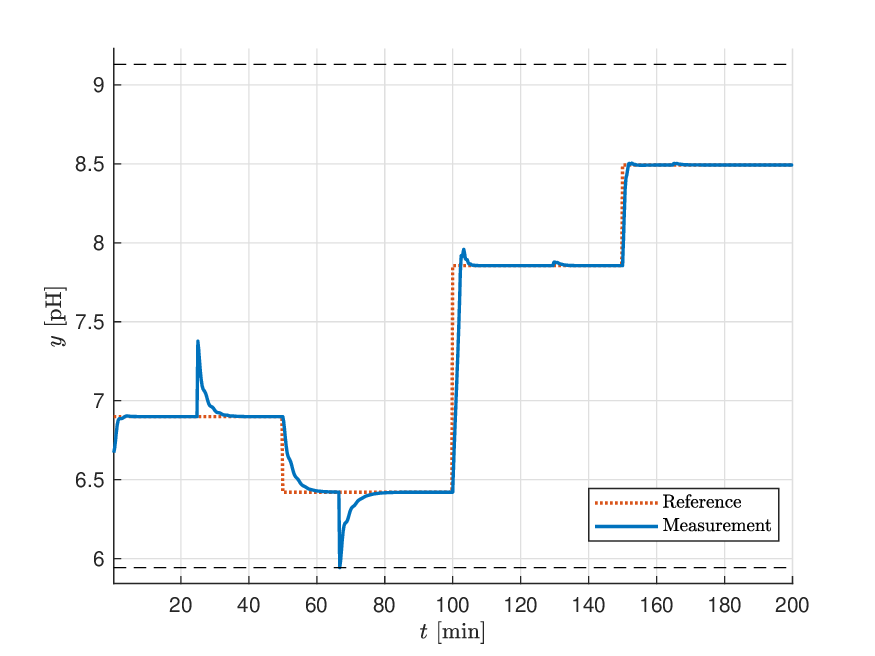}%{MPC_sim_output.eps}
    \caption{Closed-loop output performance. Black dashed lines denote output constraints.}
      \label{fig:output_cl}
\end{figure}

\begin{figure}%[htbp]
     \centering
    \includegraphics[width=0.9\columnwidth]{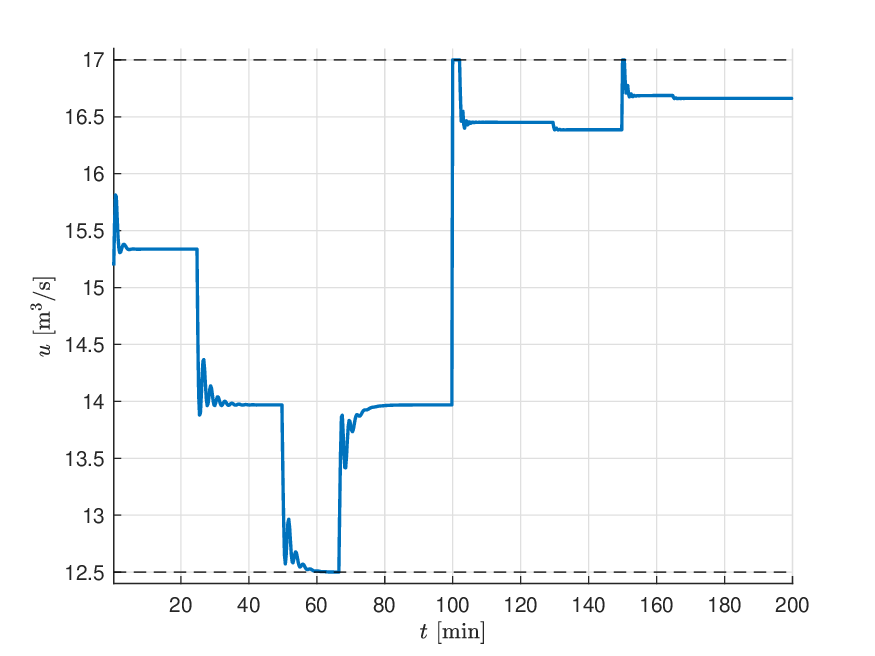}
    \caption{Evolution of the control input. Black dashed lines denote input constraints.}
      \label{fig:input_cl}
\end{figure}

In this section, the pH-neutralisation process \cite{317975} is employed to validate the theoretical results.\smallskip\\
The plant is composed of two tanks. Tank $1$, which serves as the reactor, is fed by three inputs: the inflow rate $q_{1\mathrm{e}}$, the buffer flow rate $q_2$, and the alkaline base flow rate $q_3$. The flow rate $q_{1\mathrm{e}}$ is obtained by feeding Tank $2$ with an acid flow rate $q_1$. Since Tank $2$ is characterised by significantly faster dynamics than the others involved, it is assumed that $q_{1\mathrm{e}} = q_1$. The flow rate $q_3$ is modulated by a controllable valve, while flow rates $q_{1}$ and $q_2$ are non-controllable, and are assumed to be fixed at their nominal values. The pH of the output flow rate of Tank $1$, i.e. $q_4$, is measured. \\
The overall model is a nonlinear single-input single-output system, with the controllable input defined as $u = q_3$ and the measured output as $y = \text{pH}(q_4)$. Both variables are subject to saturation constraints, namely $u \in [12.5,\ 17]$ and $y \in [5.94,\ 9.13]$. A detailed description of the process and its parameters is provided in \cite{317975}.\\
To implement the proposed control algorithm, an input–output dataset has been collected under nominal operating conditions with a sampling time of $15$~s, by exciting the simulator with a multilevel pseudo-random signal designed to cover different operating regions. The dataset has been subsequently normalised so that the input and output constraints correspond to $u \in [0,1]$ and $y \in [0,1]$, respectively. Based on the normalised data, an RNN-based model of the class \eqref{eq:plant_dynamics}, with $n=7$ states, and with $\sigma_i=\tanh(\cdot)$ for $i=1,\dots,\nu$, where $\nu=3$, has been identified using the approach in~\cite{RAVASIO2027106658}.\\
To assess the offset-free tracking capabilities, the proposed MPC-based control scheme is applied to the pH-neutralisation simulator. The control objective is to track a piecewise constant reference signal in the presence of modelling uncertainties and unknown disturbances. In particular, the following disturbances are applied to the system to test the controller robustness.  A constant additive disturbance on the plant output with amplitude $d_y = 0.15 \,[\text{pH}]$ is applied over the interval $t \in [24.5,\,66.5] \,[\text{min}]$. In addition, the input flow rate $q_3$ is changed from the nominal value of $0.55 \,[\text{m}^3/\text{s}]$ to $0.88 \,[\text{m}^3/\text{s}]$ over the interval $t \in [128.5,\,164] \,[\text{min}]$.\\ Figures~\ref{fig:output_cl} and \ref{fig:input_cl} present the closed-loop simulation results. Figure~\ref{fig:output_cl} shows that the system output successfully tracks the assigned setpoint, achieving zero tracking error asymptotically despite the presence of disturbances and modelling uncertainties. Furthermore, Figures~\ref{fig:output_cl} and \ref{fig:input_cl} demonstrate that both the input and the output remain within the prescribed constraints throughout the simulation.
%%%%%%%%%%%%%%%%%%%%%%%%%%%%%%%
% \begin{figure}%[htbp]
%      \centering
%     \fontsize{8}{12}\selectfont
%     \centering
%     \vspace{80pt}
%     \def\svgwidth{0.7\columnwidth}
%     \input{ph_neutralization}
%     \caption{pH-neutralization process.}
%     \label{fig:ph_neutralization}
% \end{figure}
%
%
%%%%%%%%%%%%%%%%%%%%%%%%%%%%%%%%%%%%%%%%%%%%%%%%%%%%%%%%%%%%%%%%%%%%%%%%%
%
%
\section{Conclusions}\label{sec:conclusions}
In this paper, the velocity form approach has been extended to a class of deep RNN models. Moreover, by exploiting a generalised incremental sector condition, we derived an LMI-based procedure for the design of a nonlinear control law ensuring global or regional stability of the origin and for computing an associated invariant set. These results were instrumental for the design of an offset-free NMPC that uses the velocity form model as the prediction model and the static control law and invariant set as terminal ingredients.
To address the general case in which the system state is not measurable, we derived LMI-based conditions for the design of a state observer. Future research will focus on developing an observer that estimates the velocity form state dynamics directly, thus removing the need for an explicit disturbance model.
%
%
%%%%%%%%%%%%%%%%%%%%%%%%%%%%%%%%%%%%%%%%%%%%%%%%%%%%%%%%%%%%%%%%%%%%%%%%%
%
%
\appendix
\section{Proof of the main results}

In this appendix we report the proofs of the main results presented in the paper. For the sake of conciseness, time dependencies are omitted where possible.\smallskip\\
%Before proceeding with the proofs, we introduce the following lemma.
%
%
%
%
%
%
\textbf{Proof of Lemma~\ref{lem:Hurwitz_D_stability}}
First, note that $B_s \in \mathbb{B}_\Theta$ if and only if, for any matrix $\Theta \in \mathbb{D}_+^{n}$ such that $\Theta \preceq I_n$, it holds that $\det(I_n - \Theta B_s) \neq 0$, i.e., $\Theta B_s$ does not have unitary eigenvalues.\\
Left- and right-multiplying \eqref{eq:digonal_hurwitz_stability} by $Q=P^{-1}$, we can write
\begin{equation*}
QB_s^\top + B_sQ - 2Q \prec 0\,.
%\Theta E P\Theta + \Theta P E^\top\Theta - 2\Theta P\Theta \prec 0\,.
\end{equation*}
Noting that $Q\succ0$ and left- and right-multiplying this inequality by $\Theta$, we obtain
\begin{equation*}
\Theta (QB_s^\top + B_sQ - 2Q)\Theta \prec 0\,.
%\Theta E P\Theta + \Theta P E^\top\Theta - 2\Theta P\Theta \prec 0\,.
\end{equation*}
Define $\tilde{Q} \coloneq Q \Theta$; thus, the above inequality can be rewritten as
\begin{equation*}
\tilde Q (B_s^\top\Theta)  +  (\Theta B_s)\tilde Q - 2\Theta \tilde Q \prec 0\,.
\end{equation*}
Since $\Theta, Q \in \mathbb{D}_+^n$, it follows that $\tilde{Q} \in \mathbb{D}_+^n$.
Moreover, in view of the fact that $\Theta \tilde{Q} \preceq \tilde{Q}$, it holds that
\begin{equation*}
 (\Theta B_s)\tilde Q + \tilde Q(\Theta B_s)^\top  - 2 \tilde Q \prec 0\,,
\end{equation*}
which, according to~\cite{boyd1994linear}, is a sufficient condition to guarantee that $\Re(\lambda_{\text{max}}(\Theta B_s)) \prec 1$, concluding the proof.
\hfill{}$\square$\smallskip\\
\textbf{Proof of Lemma~\ref{lem:sector_condition}.}
Consider the nonlinear function $q_i(v_i)=v_i-\sigma_i(v_i)$ where $\sigma_i(\cdot)$ satisfies Assumption~\ref{ass:sigmoid_function}. Also, define the function 
\[\Delta q_i\coloneq\Delta q_i(v_i,v_i+\Delta v_i)=q_i(v_i+ \Delta v_i)-q_i(v_i).\]  Then, in view of \cite[Lemma~2]{ravasio2025recurrentneuralnetworkbasedrobust}, for all $h_i>1$, $\exists\bar v_i(h_i)\in\R_{+}$ such that, for all pairs $(v_i,v_i+\Delta v_i)\in[-\bar v_i(h_i),\bar v_i(h_i)]^2$, it holds that
\begin{equation}\label{eq:sector_condition_q}
\Delta q_i(\Delta v_i-h_i\Delta q_i)\geq 0,
\end{equation}
Function $\bar v_i(h_i):(1,+\infty)\to(0,+\infty)$ is a continuous, strictly decreasing function such that $\bar v_i(h_i) \to +\infty$ as $h_i\to 1^+$ and $\bar v_i(h_i)\to0$ as $h_i\to +\infty$. Also, in case $h_i=1$, condition \eqref{eq:sector_condition_q} holds for all $(v_i,v_i+\Delta v_i)\in\R^2$.\\
We now introduce the parameter $\lambda_i=1-1/h_i\in[0,1)$. Noting that $\Delta q_i=\Delta v_i-\Delta s_i$, and multiplying both sides of \eqref{eq:sector_condition_q} by $1/h_i$, we obtain
\[(\Delta v_i-\Delta s_i)\left(\left(\frac{1}{h_i}-1\right)\Delta v_i+\Delta s_i\right)\geq0,\]
which is equivalent to condition \eqref{eq:sector_condition_s} after substituting $h_i=1/(1-\lambda_i)$. We now parametrise the function $\bar v_i(h_i)$ with respect to $\lambda_i$ by defining
\[\bar v_i(\lambda_i)\coloneq\bar v_i(h_i{=}1/(1-\lambda_i)).\]
The funciton $\bar v_i(\lambda_i):(0,1)\to(0,+\infty)$ is continuous and strictly decreasing, and such that $\bar v_i(\lambda_i)\to+\infty$ as $\lambda_i\to0^+$ and $\bar v_i(\lambda_i)\to0$ as $\lambda_i\to1^-$.
\hfill{}$\square$\smallskip\\
\textbf{Proof of Proposition~\ref{prop:mapping_xi_xu}}
The proof of Proposition~\ref{prop:mapping_xi_xu} is organised in three steps, specified here for better clarity.
\begin{itemize}
\item[1.] Derive \eqref{eq:mapping_xi_xu}.
\item[2.] Show that under Assumptions~\ref{ass:sigmoid_function} and \ref{ass:velocity_form_ass}-\emph{(i)}, the implicit nonlinear equation \eqref{eq:nonlinear_function} admits a unique solution.%, i.e., the relation \eqref{eq:mapping_xi_xu} is well defined.
\item[3.] Conclude that, if also Assumption~\ref{ass:velocity_form_ass}-\emph{(ii)}, for all $\bar y$, the mapping $(x(k-1), u(k-1)) \mapsto \xi(k)$ in \eqref{eq:mapping_xi_xu} is bijective.
\end{itemize}
1. To derive \eqref{eq:mapping_xi_xu}, we rewrite system~\eqref{eq:plant_dynamics} as  
\begin{equation*}
\begin{cases*}
    x = A x^- + B u^- + B_s s^- \\
    y = CA x^- + CB u^- + CB_s s^-
\end{cases*}\,.
\end{equation*}
By subtracting $x^-$ from both sides of the first equation and $\bar y$ from both sides of the second and recalling that $\Delta x = x - x^-$ and $\epsilon = y - \bar y$, we obtain  
\begin{equation}\label{eq:mapping_xi_xu_ext}
\begin{cases}
\Delta x = (A - I_n) x^- + B u^- + B_s s^- \\
\epsilon = C A x^- + C B u^- + C B_s s^- - \bar y
\end{cases}\,,
\end{equation}
which is equivalent to $\eqref{eq:mapping_xi_xu}$.\smallskip\\
%
%
%To prove that the mapping $ (x^-, u^-, \bar{y}) \mapsto \xi $ is bijective, we first show that \eqref{eq:nonlinear_function} admits a unique solution.\\
2. According to Dini's Implicit Function Theorem~\cite{Apostol1974}, a sufficient condition for the existence of $s = f_s(x, u)$ satisfying~\eqref{eq:nonlinear_function} is the invertibility of the Jacobian
\begin{align*}
J_s &= \frac{\partial}{\partial s} \left( s - \sigma(\tilde{A}x + \tilde{B}u + \tilde{B}_s s) \right) \\
&= I_\nu - \frac{\partial \sigma(\tilde{A}x + \tilde{B}u + \tilde{B}_s s)}{\partial s} \\
%&= I_\nu - \frac{\partial \sigma(v)}{\partial v} \tilde{B}_s\\
&= I_\nu - \diag\left(\frac{\partial \sigma_1(v_1)}{\partial v_1},\dots,\frac{\partial \sigma_\nu(v_\nu)}{\partial v_\nu}\right) \tilde{B}_s\,,
\end{align*}
where $v_i = \tilde{A}_ix + \tilde{B}_iu + \tilde{B}_{s,i} s $, for $i=1,\dots,\nu$.  
Noting that, under Assumption~\ref{ass:sigmoid_function}, 
\(\frac{\partial \sigma_i(v_i)}{\partial v_i} \in (0, 1] \)
for $i = 1, \dots, \nu $, it follows that $J_s$ is invertible by Assumption~\ref{ass:velocity_form_ass}-\emph{(i)}.\smallskip\\
3. Define the map $F(x^-,u^-,\xi)=C_\xi\phi(x^-,u^-)+C_{\bar y}\bar y-\xi$. According to Dini’s Implicit Function Theorem, a sufficient condition for the existence of a unique pair $(x,u)$ corresponding to a given $\xi$ satisfying \eqref{eq:mapping_xi_xu}, i.e. such that  $F(x^-,u^-,\xi)=0$, is the invertibility of the Jacobian,
\begin{equation}\label{eq:jacobian_xi}
    \begin{aligned}
        J_\xi&=\cfrac{\partial F(x^-,u^-,\xi)}{\partial(x^-,u^-)}\\
        &=C_\xi\begin{bmatrix}
            I&0\\0&I\\\cfrac{\partial f_\mathrm s(x^-,u^-)}{\partial x^-}&\cfrac{\partial f_\mathrm s(x^-,u^-)}{\partial u^-}
        \end{bmatrix}.
    \end{aligned}
\end{equation}
Using~\eqref{eq:nonlinear_function}, and defining $\Theta=\diag(\theta_1,\dots,\theta_\nu)$, where $\theta_i=\partial\sigma(v_i^-)/\partial v_i^-$, for $i=1,\dots,\nu$, it holds
\begin{align*}
\cfrac{\partial f_\mathrm s(x^-,u^-)}{\partial x^-}&=\Theta\frac{\partial v^-}{\partial x^-}
\\&=\Theta\left(\tilde A+\tilde B_s\frac{\partial f_s(x^-,u^-)}{\partial x^-}\right).
\end{align*}
Since $I_\nu-\Theta\tilde B_s$ is invertible by Assumption~\ref{ass:velocity_form_ass}-\emph{(i)}, it follows that
\begin{equation*}
\cfrac{\partial f_\mathrm s(x^-,u^-)}{\partial x^-}=(I_\nu-\Theta\tilde B_s)^{-1}
\Theta \tilde A.
\end{equation*}
Applying a similar reasoning, it is possible to show that 
\begin{equation*}
\cfrac{\partial f_\mathrm s(x^-,u^-)}{\partial u^-}=(I_\nu-\Theta\tilde B_s)^{-1}
\Theta \tilde B.
\end{equation*}
Substituting these expressions into \eqref{eq:jacobian_xi}, we obtain 
\[J_\xi=\begin{bmatrix} A - I_n+B_s\Phi^{-1}\Theta\tilde A & B+B_s\Phi^{-1}\Theta\tilde B \\ C(A+B_s\Phi^{-1}\Theta\tilde A) 
    & C(B+B_s\Phi^{-1}\Theta\tilde B)
    \end{bmatrix}.\] 
The inverse of $J_\xi$ always exists in view of Assumption~\ref{ass:velocity_form_ass}-\emph{(ii)}.
\smallskip\\
%3. In view of Assumption~\ref{ass:sigmoid_function}, it holds that $ v_i \sigma_i(v_i) \geq 0 $ and $ |\sigma_i(v_i)| \leq |v_i| $ for $ i = 1, \dots, \nu $. Therefore, for each $ i $, there exists $ \theta_i \in (0, 1] $ such that $ s_i = \sigma_i(v_i) = \theta_i v_i $.  Defining $ \Theta = \diag(\theta_1, \dots, \theta_\nu) $, it follows that
%\[
%s = \Theta (\tilde{A}x + \tilde{B}u + \tilde{B}_s s).
%\]
%In view of Assumption~\ref{ass:velocity_form_ass}-\emph{(i)}, leads to the solution
%\[
%s = (I_\nu - \Theta \tilde{B}_s)^{-1} \Theta(\tilde{A}x + \tilde{B}u).
%\]
%Substituting this expression into \eqref{eq:mapping_xi_xu_ext}, we obtain
%\[
%\xi = M 
%\begin{bmatrix}
%    x^- \\ u^-
%\end{bmatrix}
%+ C_{\bar{y}} \bar{y}.
%\]
%Note that $ M^{-1} $ exists by Assumption~\ref{ass:velocity_form_ass}-\emph{(ii)}. Hence, we can compute $ (x^-, u^-) $ as
%\[
%\begin{bmatrix}
%    x^- \\ u^-
%\end{bmatrix}
%= M^{-1} (\xi - C_{\bar{y}} \bar{y}),
%\]
%which concludes the proof.
%\hfill{}$\square$\smallskip\\
%
%
%
%
%
%
%
\textbf{Proof of Theorem \ref{th:asy_stability_ctr}.}
The proof of Theorem~\ref{th:asy_stability_ctr} is divided in three steps, specified here for better clarity.
\begin{itemize}
\item[1.] %For $i=1,\dots,\nu$ define $v_i(k)=\tilde A_ix(k)+\tilde B_iu(k)+\tilde B_{s,i}s(k)$ and $\Delta v(k) = v(k)-v(k-1)$, where $v(k)=[v_1(k),\dots, v_\nu(k)]^\top$. 
Show that if \eqref{eq:local_stability_LMI} holds and if $v(k-1),v(k)\in\mathcal V(\Lambda_\mathrm c)$, where $v(k)\coloneq \tilde Ax(k)+\tilde Bu(k)+\tilde B_{s}s(k)$, then 
\begin{equation}\label{eq:asy_stab_lyapunov}
V_\mathrm c(k+1)-V_\mathrm c(k)<0\,,
\end{equation}
where $V_\mathrm c(k)\coloneq \norm{\xi(k)}_{P_\mathrm c}^2$;
\item[2.] Show that, under Assumption~\ref{ass:velocity_form_ass}, the condition $\xi(k)\in\mathcal E(P_\mathrm c/\gamma)$ implies $\xi(k+1)\in\mathcal E(P_\mathrm c/\gamma)$ for all $\gamma\in\R_+$ if $\Lambda_\mathrm c=0$ and for $\gamma\in(0,\gamma_\mathrm c]$ if $\Lambda_\mathrm c\neq0$;
\item[3.] Show that \eqref{eq:condition_implicit_function_ctr} implies that \eqref{eq:control_law} is well-defined.
\end{itemize}
1. In view of Lemma~\ref{lem:sector_condition} and \eqref{eq:sector_inequality}, and noting that $\Delta s_\mathrm c=\Delta s(v^-,v)$, for any $\Lambda_\mathrm c\in\mathbb{D}_{\geq0}^\nu$, such that $\Lambda_\mathrm c\prec I_\nu$, if $v^-,v\in\mathcal{V}(\Lambda_\mathrm c)$, then for all $S_\mathrm c\in\mathbb{D}_+^\nu$,
\begin{equation}\label{eq:sector_inequality_ctr}
    (\Delta v-\Delta s_\mathrm c)^\top S_\mathrm c(\Delta s_\mathrm c-\Lambda_\mathrm c\Delta v)\geq 0,
\end{equation}
where $\Delta v = v-v^-$.
Moreover, using \eqref{eq:control_law} it holds that
\begin{align*}
\Delta v&=%v-v^-\\&=
\tilde{A}\Delta x+\tilde B\Delta u+\tilde B_s\Delta s_\mathrm c\\
&=\tilde{\mathcal{A}}\xi+\tilde B\Delta u+\tilde B_s\Delta s_\mathrm c\\
&=\tilde{\mathcal{A}}_{\mathrm K}\xi+\mathcal{\tilde{B}}_{s,\mathrm K}\Delta s_\mathrm c\,.
\end{align*}
Therefore, condition \eqref{eq:sector_inequality_ctr} can be rewritten as
\begin{multline}\label{eq:sector_inequality_ctr_extended_0}
    \left(\tilde{\mathcal{A}}_{\mathrm K}\xi+(\mathcal{\tilde{B}}_{s,\mathrm K}-I_\nu)\Delta s_\mathrm c\right)^\top S_\mathrm c\Big(-\Lambda_\mathrm c\tilde{\mathcal{A}}_{\mathrm K}\xi+\\+(I_\nu-\Lambda_\mathrm c\mathcal{\tilde{B}}_{s,\mathrm K})\Delta s_\mathrm c\Big)\geq 0.
\end{multline}
Define $\phi_{\mathrm c}=[\xi^\top,\ \Delta s_\mathrm c^\top]^\top$. Condition \eqref{eq:sector_inequality_ctr_extended_0} implies that
\begin{multline*}
    \phi_{\mathrm c}^\top\begin{bmatrix}\tilde{\mathcal{A}}_{\mathrm K}^\top\\\mathcal{\tilde{B}}_{s,\mathrm K}^\top-I_\nu\end{bmatrix}S_\mathrm c\begin{bmatrix}-\Lambda_\mathrm c\tilde{\mathcal{A}}_{\mathrm K}&I_\nu-\Lambda_\mathrm c\mathcal{\tilde{B}}_{s,\mathrm K}\end{bmatrix}\phi_{\mathrm c} +\\
    \phi_{\mathrm c}^\top\begin{bmatrix}-\tilde{\mathcal{A}}_{\mathrm K}^\top\Lambda_\mathrm c\\I_\nu-\tilde{\mathcal{B}}_{s,\mathrm K}^\top\Lambda_\mathrm c\end{bmatrix}S_\mathrm c\begin{bmatrix}\tilde{\mathcal{A}}_{\mathrm K}&\mathcal{\tilde{B}}_{s,\mathrm K}-I_\nu\end{bmatrix}\phi_{\mathrm c}\geq 0,
\end{multline*}
which leads to
\begin{equation}\label{eq:sector_inequality_ctr_extended_1}
\phi_{\mathrm c}^\top\begin{bmatrix}-2\tilde{\mathcal{A}}_{\mathrm K}^\top S_\mathrm c\Lambda_\mathrm c\tilde{\mathcal{A}}_{\mathrm K}&B_{\Lambda,\mathrm c}^\top\\    B_{\Lambda,\mathrm c}&S_{\Lambda,\mathrm c}\end{bmatrix}\phi_{\mathrm c}\geq 0\,,
\end{equation}
where
\begin{align*}
    B_{\Lambda,\mathrm c}&=(I_\nu-\mathcal{\tilde{B}}_{s,\mathrm K}^\top )S_\mathrm c\Lambda_\mathrm c\tilde{\mathcal{A}}_{\mathrm K}+(I_\nu-\mathcal{\tilde{B}}_{s,\mathrm K}^\top\Lambda_\mathrm c )S_\mathrm c\tilde{\mathcal{A}}_{\mathrm K}\\
    &=S_\mathrm c\tilde{\mathcal{A}}_{\mathrm K}+(I_\nu-\mathcal{\tilde{B}}_{s,\mathrm K}^\top) S_\mathrm c\Lambda_\mathrm c\tilde{\mathcal{A}}_{\mathrm K}-\mathcal{\tilde{B}}_{s,\mathrm K}^\top S_\mathrm c\Lambda_\mathrm c\tilde{\mathcal{A}}_{\mathrm K}\,,
\end{align*}
and
\begin{align*}
S_{\Lambda,\mathrm c}&=(\mathcal{\tilde{B}}_{s,\mathrm K}^\top-I_\nu) S_\mathrm c(I_\nu-\Lambda_\mathrm c\mathcal{\tilde{B}}_{s,\mathrm K})+\\&\quad+(I_\nu-\mathcal{\tilde{B}}_{s,\mathrm K}^\top\Lambda_\mathrm c) S_\mathrm c(\mathcal{\tilde{B}}_{s,\mathrm K}-I_\nu)\\
&=(\mathcal{\tilde{B}}_{s,\mathrm K}-I_\nu)^\top S_\mathrm c+S_\mathrm c(\mathcal{\tilde{B}}_{s,\mathrm K}-I_\nu)+\\
&\quad+(I_\nu-\mathcal{\tilde{B}}_{s,\mathrm K})^\top S_\mathrm c\Lambda_\mathrm c\mathcal{\tilde{B}}_{s,\mathrm K}+\mathcal{\tilde{B}}_{s,\mathrm K}^\top S_\mathrm c\Lambda_\mathrm c(I_\nu-\mathcal{\tilde{B}}_{s,\mathrm K})\,.
\end{align*}
By separating the terms that depend on $\Lambda_\mathrm c$ in \eqref{eq:sector_inequality_ctr_extended_1}, we obtain
\begin{equation}\label{eq:sector_inequality_ctr_extended}
\phi_\mathrm c^\top
\Bigg(\begin{bmatrix}0&\tilde{\mathcal{A}}_{\mathrm K}^\top S_\mathrm c\\
    S_\mathrm c\tilde{\mathcal{A}}_{\mathrm K}&-U_{S,\mathrm c}\end{bmatrix}-M_{\mathrm c}(\Lambda_\mathrm c)\Bigg)\phi_\mathrm c\geq 0\,.\end{equation}
%where
%\begin{align*}
%    &M_{\Gamma,\mathrm c}=\begin{bmatrix}2\tilde{\mathcal{A}}_{\mathrm K}^\top S_\mathrm c\Gamma_c\tilde{\mathcal{A}}_{\mathrm K}\!\!\!\!&\!\!\!\!\tilde{\mathcal{A}}_{\mathrm K}^\top S_\mathrm c\Gamma_c(2\mathcal{\tilde{B}}_{s,\mathrm K}-I_\nu)\\
%    \star\!\!\!\!&\!\!\!\!(\mathcal{\tilde{B}}_{s,\mathrm K}-I_\nu)^\top S_\mathrm c\Gamma_c\mathcal{\tilde{B}}_{s,\mathrm K}+\mathcal{\tilde{B}}_{s,\mathrm K}^\top \Gamma_cS_\mathrm c(\mathcal{\tilde{B}}_{s,\mathrm K}-I_\nu)\end{bmatrix}\\
%    &=\begin{bmatrix}
%        \tilde{\mathcal{A}}_{\mathrm K}^\top & \tilde{\mathcal{A}}_{\mathrm K}^\top\\\mathcal{\tilde{B}}_{s,\mathrm K}^\top & (\mathcal{\tilde{B}}_{s,\mathrm K}-I_\nu)^\top
%    \end{bmatrix}
%    \begin{bmatrix}
%        S_\mathrm c\Gamma_c & 0\\
%        0 & S_\mathrm c\Gamma_c
%    \end{bmatrix}
%    \begin{bmatrix}
%        \tilde{\mathcal{A}}_{\mathrm K} & \mathcal{\tilde{B}}_{s,\mathrm K}\\\tilde{\mathcal{A}}_{\mathrm K} & \mathcal{\tilde{B}}_{s,\mathrm K}-I_\nu
%    \end{bmatrix}.
%\end{align*}
%
Now, using \eqref{eq:closed_loop_velocityform}, we can write \begin{equation}\label{eq:deltaV_c}
    \begin{aligned}
        \Delta V_\mathrm c&\!\!=V_\mathrm c(k+1)-V_\mathrm c(k)\\
        &\!\!=\xi(k+1)^\top P_\mathrm c\xi(k+1)-\xi(k)^\top P_\mathrm c\xi(k)\\
        &\!\!=\!\phi_\mathrm c^\top\left(\begin{bmatrix}
            \mathcal{A}_\mathrm K^\top\\\mathcal{B}_{s,\mathrm K}^\top
        \end{bmatrix}P_\mathrm c\begin{bmatrix}
            \mathcal{A}_\mathrm K&\mathcal{B}_{s,\mathrm K}
        \end{bmatrix}{-}\begin{bmatrix}
            P_\mathrm c &0\\0&0
        \end{bmatrix}\right)\phi_\mathrm c.
    \end{aligned}
    \end{equation}
We can exploit \eqref{eq:sector_inequality_ctr_extended} to guarantee $\Delta V_\mathrm c<0$, and therefore that the origin is an  asymptotically stable equilibrium for \eqref{eq:closed_loop_velocityform},
by imposing
\begin{multline}\label{eq:condition_local_stability}
    \Delta V_\mathrm c+\phi_\mathrm c^\top\left(\begin{bmatrix}0&\tilde{\mathcal{A}}_{\mathrm K}^\top S_\mathrm c\\
    S_\mathrm c\tilde{\mathcal{A}}_{\mathrm K}&-U_{S,\mathrm c}\end{bmatrix}-M_{\mathrm c}(\Lambda_\mathrm c)\right)\phi_\mathrm c<0,\\ \forall v^-,v\in\mathcal{V}(\Lambda_\mathrm c).
\end{multline}
Substituting \eqref{eq:deltaV_c} in \eqref{eq:condition_local_stability} leads to
\begin{multline*}
\phi_\mathrm c^\top\Bigg(\begin{bmatrix}P_\mathrm c&-\tilde{\mathcal{A}}_{\mathrm K}^\top S_\mathrm c\\
    -S_\mathrm c\tilde{\mathcal{A}}_{\mathrm K}&U_{\mathrm S,\mathrm c}\end{bmatrix}+\\-\begin{bmatrix}
            \mathcal{A}_\mathrm K^\top\\\mathcal{B}_{s,\mathrm K}^\top
        \end{bmatrix}P_\mathrm c\begin{bmatrix}
            \mathcal{A}_\mathrm K&\mathcal{B}_{s,\mathrm K}
        \end{bmatrix}+M_{\mathrm c}(\Lambda_\mathrm c)\Bigg)\phi_\mathrm c>0,
\end{multline*}
which is satisfied for all $v^-,v\in\mathcal{V}(\Lambda_\mathrm c)$ if \eqref{eq:local_stability_LMI} holds.\smallskip\\
2. In case $\mathcal{I}(\Lambda_\mathrm c) = \emptyset$, condition \eqref{eq:local_stability_LMI} implies that \eqref{eq:asy_stab_lyapunov} holds for all $(x, u) \in \mathbb{R}^n \times \mathbb{R}^m$. Therefore, for any $\gamma \in \mathbb{R}_+$, the condition $\xi \in \mathcal{E}(P_\mathrm c / \gamma)$ implies that 
\(V_\mathrm c(k+1) < V_\mathrm c(k) \leq \gamma\,,\) 
i.e., $\xi^+ \in \mathcal{E}(P_\mathrm c / \gamma)$.\\
In case $\mathcal{I}(\Lambda_\mathrm c) \neq \emptyset$, condition \eqref{eq:local_stability_LMI} implies that \eqref{eq:asy_stab_lyapunov} holds for all $v^-, v \in \mathcal{V}(\Lambda_\mathrm c)$. To address this case, we need to show that there exists $\gamma_\mathrm c \in \mathbb{R}_+$ such that, for all $\gamma \in (0, \gamma_\mathrm c]$, if $\xi \in \mathcal{E}(P_\mathrm c / \gamma)$, then $v^-, v \in \mathcal{V}(\Lambda_\mathrm c)$.\\
In view of Proposition~\ref{prop:mapping_xi_xu}, $\xi\in\mathcal E(P_\mathrm c/\gamma)$ if and only if 
$\phi(x^-,u^-)\in\{\phi\in\R^{n_\xi+\nu}: (C_\xi\phi+C_{\bar y}\bar y)^\top P_\mathrm c(C_\xi\phi+C_{\bar y}\bar y)\leq\gamma\}$.\\
Therefore, by the definition of $\gamma_\mathrm c$ in \eqref{eq:PI_set_gamma_ctr}, it is guaranteed that for all $\gamma\in(0,\gamma_\mathrm c]$ if $\xi \in \mathcal{E}(P_\mathrm c / \gamma)$, then $G_\mathrm c\phi(x^-,u^-)\leq \bar b_\mathrm c$, i.e., $|v_i^-|\leq \bar v_i(\lambda_{\mathrm c,i})$ for all $i\in \mathcal I(\Lambda_\mathrm c)$. 
Moreover, in view of Proposition~\ref{prop:mapping_xi_xu},
\[\xi^+=C_\xi\phi(x,u)+C_{\bar y}\bar y\,.\]
Therefore, the same arguments shows that if $\xi^+ \in \mathcal{E}(P_\mathrm c/\gamma_\mathrm c)$, then $G_\mathrm c \phi(x, u) \leq \bar{b}_\mathrm c$, implying $v \in \mathcal{V}(\Lambda_\mathrm c)$.
Since~\eqref{eq:asy_stab_lyapunov} holds for all $v^-, v \in \mathcal{V}(\Lambda_\mathrm c)$, we can conclude that $\xi\in\mathcal{E}(P_\mathrm c/\gamma)$ implies $\xi^+\in\mathcal{E}(P_\mathrm c/\gamma)$, %, the same argument shows that $|v_i|\leq \bar v_i(\lambda_{\mathrm c,i})$ for all $i\in \mathcal I(\Lambda_\mathrm c)$, 
completing the proof.
\smallskip\\
3. First, note that condition \eqref{eq:condition_implicit_function_ctr} can be rewritten as
\[
\tilde{\mathcal{B}}_{s,\mathrm K}^\top S_\mathrm c + S_\mathrm c \tilde{\mathcal{B}}_{s,\mathrm K} - 2S_\mathrm c \prec 0,
\]
which, recalling that $S_\mathrm c \in \mathbb{D}_+^\nu$ and applying Lemma~\ref{lem:Hurwitz_D_stability}, implies that $\tilde{\mathcal{B}}_{s,\mathrm K} \in \mathbb{B}_\Theta$.
\\
Now, since $\Delta s_\mathrm c = s(v) - s(v^-)$ and $v = \Delta v + v^-$, it follows that $\Delta s$ in \eqref{eq:control_law} is the solution to the equation
\begin{equation}\label{eq:implicit_function_Deltas}
\Delta s_\mathrm c-\sigma(\tilde{\mathcal{A}}_{\mathrm e,\mathrm K}\xi+\tilde{\mathcal{B}}_{s,\mathrm K}\Delta s_\mathrm c+v^-)+\sigma(v^-)=0\,.
\end{equation}
%Before proving the existence of a solution to \eqref{eq:implicit_function_Deltas}, note that condition \eqref{eq:condition_invertibility_deltas} implies that matrix $\tilde{\mathcal{B}}_{s,\mathrm K}-I_\nu$ is diagonally stable. As a consequence, $\Re(\lambda_\text{min}(I_\nu-\tilde{\mathcal{B}}_{s,\mathrm K}))>0$. Therefore, for all  $\Theta\in\mathbb D_+^\nu$ such that $\Theta\preceq I_\nu$, it holds that $\Re(\lambda_{\text{min}}(I_\nu-\Theta\tilde{\mathcal{B}}_{s,\mathrm K}))>0$, and hence $\rank(I-\Theta\tilde{\mathcal{B}}_{s,\mathrm K})=\nu$.\\
In view of Dini's Implicit Function Theorem, a sufficient condition for \eqref{eq:implicit_function_Deltas} to admit a unique solution is the invertibility of the Jacobian
\begin{align*}
J_{\Delta s_\mathrm c}&{=}\cfrac{\partial}{\partial (\Delta s_\mathrm c)}\left(\Delta s_\mathrm c{-}\sigma(\tilde{\mathcal{A}}_{\mathrm e,\mathrm K}\xi{+}\tilde{\mathcal{B}}_{s,\mathrm K}\Delta s_\mathrm c+v^-)+\sigma(v^-)\right)\\
&=I_\nu-\cfrac{\partial}{\partial v}\left( \sigma(\tilde{\mathcal{A}}_{\mathrm e,\mathrm K}\xi+\tilde{\mathcal{B}}_{s,\mathrm K}\Delta s_\mathrm c+v^-)\right)\tilde{\mathcal{B}}_{s,\mathrm K}\\
&=I_\nu-\diag\left(\cfrac{\partial \sigma_1(v_1)}{\partial v_1},\ \dots,\  \cfrac{\partial \sigma_1(v_\nu)}{\partial v_\nu}\right)\tilde{\mathcal{B}}_{s,\mathrm K}\,,
\end{align*}
which is verified due to the fact that $\tilde{\mathcal{B}}_{s,\mathrm K} \in \mathbb{B}_\Theta$.
\hfill{}$\square$\smallskip\\
\textbf{Proof of Theorem~\ref{th:convergence_MPC}.}
The proof of Theorem~\ref{th:convergence_MPC} resorts to standard MPC arguments. Specifically, we verify that there exists a control law $\kappa(\cdot)$ such that, if $\Delta u =  \kappa(\xi(k))$ and $\xi(k) \in \mathbb{E}_\mathrm f$, then
\begin{itemize}
    \item[\textbf{c1}.] the terminal cost satisfies the condition   \begin{equation}\label{eq:terminal_cost_condition}
        \Delta V_\mathrm f\leq-\norm{\xi(k)}_Q^2-\norm{\kappa(\xi(k))}_R^2\,,
    \end{equation}
    where $\Delta V_\mathrm f\coloneq V_\mathrm f(\xi(k+1))-V_\mathrm f(\xi(k))$;
    \item[\textbf{c2}.] the state of \eqref{eq:velocityform_dynamics_compact} remains in the terminal set at the next time step, i.e., $\xi(k+1) \in \mathbb{E}_\mathrm f$;
     \item[\textbf{c3}.] the input and output constraints are satisfied, i.e., $u \in \mathbb{U}$ and $y \in \mathbb{Y}$.
\end{itemize}
First, note that if we set $\kappa_\mathrm f(\xi) = K\xi + \tilde{K} \Delta s_\mathrm c$, the closed-loop dynamics is given by~\eqref{eq:closed_loop_velocityform}. Therefore, by defining $\phi_\mathrm f = [\xi^\top,\ \Delta s_\mathrm c^\top]^\top$, it holds that
\begin{equation*}
    \Delta V_\mathrm f=\phi_\mathrm f^\top\left(\begin{bmatrix}
            \mathcal{A}_\mathrm K^\top\\\mathcal{B}_{s,\mathrm K}^\top
        \end{bmatrix}P_\mathrm f\begin{bmatrix}
            \mathcal{A}_\mathrm K&\mathcal{B}_{s,\mathrm K}
        \end{bmatrix}-\begin{bmatrix}
            P_\mathrm f &0\\0&0
        \end{bmatrix}\right)\phi_\mathrm f\,.
\end{equation*}
Expanding the right-hand side of \eqref{eq:terminal_cost_condition}, we obtain
\begin{align*}
&\norm{\xi(k)}_Q^2+\norm{\kappa(\xi(k))}_R^2\\
&=\xi^\top Q\xi+\xi^\top K^\top R K\xi+2\xi^\top K^\top R\tilde K\Delta s_\mathrm c+ \\&\quad + \Delta s_\mathrm c^\top \tilde K^\top R \tilde K\Delta s_\mathrm c\\
&=\phi_\mathrm f^\top\begin{bmatrix}
Q+K^\top R K & K^\top R \tilde{K} \\
\tilde{K}^\top R K & \tilde{K}^\top R \tilde{K}
\end{bmatrix}\phi_\mathrm f\,.
\end{align*}
Setting $v = \tilde{A}x + \tilde{B}u + \tilde{B}_s s$, $\tilde{v} = v^-$, and $\Delta v = v - v^-$, and using similar arguments as in \eqref{eq:sector_inequality_ctr}–\eqref{eq:condition_local_stability} in the proof of Theorem~3, we can prove that condition~\eqref{eq:terminal_cost_condition} holds for all $v^-, v \in \mathcal{V}(\Lambda_\mathrm f)$ by showing that
\begin{multline*}
    \Delta V_\mathrm f+\phi_\mathrm f^\top\Bigg(\begin{bmatrix}0\!\!\!\!&\!\!\!\!\tilde{\mathcal{A}}_{\mathrm K}^\top S_\mathrm f\\
    S_\mathrm f\tilde{\mathcal{A}}_{\mathrm K}&(\mathcal{\tilde{B}}_{s,\mathrm K}-I_\nu)^\top S_\mathrm f+S_\mathrm f(\mathcal{\tilde{B}}_{s,\mathrm K}-I_\nu)\end{bmatrix}\\-M_\mathrm c(\Lambda_\mathrm f)\Bigg)\phi_\mathrm f\leq-\phi_\mathrm f^\top\begin{bmatrix}
Q+K^\top R K & K^\top R \tilde{K} \\
\tilde{K}^\top R K & \tilde{K}^\top R \tilde{K}
\end{bmatrix}\phi_\mathrm f,\\\forall v^-,v\in\mathcal{V}(\Lambda_\mathrm f)\,.
\end{multline*}
Recalling that where $U_{\mathrm S,\mathrm f}\coloneq(I_\nu - \tilde{\mathcal{B}}_{s,\mathrm K})^\top S_\mathrm f + S_\mathrm f (I_\nu - \tilde{\mathcal{B}}_{s,\mathrm K})- \tilde K^\top R\tilde K$, this last condition can be rewritten as 
\begin{multline*}
   \phi_\mathrm f^\top\Bigg(\begin{bmatrix}P_\mathrm f-Q-K^\top R K&-\tilde{\mathcal{A}}_{\mathrm K}^\top S_\mathrm f-K^\top R\tilde K \\
    -S_\mathrm f\tilde{\mathcal{A}}_{\mathrm K}-\tilde K^\top RK&U_{\mathrm S,\mathrm f}\end{bmatrix}\\-\begin{bmatrix}
            \mathcal{A}_\mathrm K^\top\\\mathcal{B}_{s,\mathrm K}^\top
        \end{bmatrix}P_\mathrm f\begin{bmatrix}
            \mathcal{A}_\mathrm K&\mathcal{B}_{s,\mathrm K}
        \end{bmatrix}+M_\mathrm c(\Lambda_\mathrm f)\Bigg)\phi_\mathrm f\geq0,\\
        \forall v^-,v\in\mathcal{V}(\Lambda_\mathrm f)\,,
\end{multline*}
which is satisfied in view of \eqref{eq:terminal_set_LMI}.\\
In the trivial case $\mathcal{I}(\Lambda_\mathrm f) = \emptyset$, condition~\textbf{c1} is satisfied for all $(x, u) \in \mathbb{R}^n \times \mathbb{R}^m$. Moreover, since~\eqref{eq:terminal_cost_condition} implies $\Delta V_\mathrm f < 0$, it follows that if $\xi(k) \in \mathbb{E}_\mathrm f$, then
\(
V_\mathrm f(\xi(k+1)) < V_\mathrm f(\xi(k)) \leq \gamma_\mathrm f\,,
\)
which in turn implies $\xi(k+1) \in \mathbb{E}_\mathrm f$.\\
In the case $\mathcal{I}(\Lambda_\mathrm f) \neq \emptyset$, condition~\eqref{eq:terminal_cost_condition} holds only if $v^-, v \in \mathcal{V}(\Lambda_\mathrm f)$. However, by construction of $\gamma_\mathrm f$ in Step~5-b, we have $G_\mathrm f \phi(x^-, u^-) \leq \bar{b}_\mathrm f$ for all $\xi \in \mathbb{E}_\mathrm f$, which implies $v^- \in \mathcal{V}(\Lambda_\mathrm f)$. Furthermore, in view of Proposition~\ref{prop:mapping_xi_xu},
%\[\xi^+=C_\xi\phi(x,u)+C_{\bar y}\bar y\,.\]
if 
$\xi^+ \in \mathbb{E}_\mathrm f$, then $G_\mathrm f \phi(x, u) \leq \bar{b}_\mathrm f$, implying $v \in \mathcal{V}(\Lambda_\mathrm f)$.
Since~\eqref{eq:terminal_cost_condition} holds for all $v^-, v \in \mathcal{V}(\Lambda_\mathrm f)$, we can conclude that $\xi \in \mathbb{E}_\mathrm f$ implies $\xi^+ \in \mathbb{E}_\mathrm f$, thus satisfying conditions~\textbf{c1} and~\textbf{c2}.\\
Finally, noting that $G_\mathrm f \phi(x, u) \leq \bar{b}_\mathrm f$ implies $(u, y) \in \mathbb{U} \times \mathbb{Y}$, condition \textbf{c3} is also satisfied, thus concluding the proof.
\hfill{}$\square$\smallskip\\
\textbf{Proof of Theorem~\ref{th:asy_stability_obs}.}
The proof of Theorem~\ref{th:asy_stability_obs} proceeds along similar lines to the proof of Theorem~\ref{th:asy_stability_ctr}.
\begin{itemize}
    \item[1.] Defining $\hat v \coloneq\tilde{A}_\mathrm e\hat{\eta}+\tilde Bu+\tilde B_s\hat s+\tilde LC_\mathrm ee$ and $v\coloneq\tilde{A}_\mathrm e\eta+\tilde Bu+\tilde B_ss$, first prove that \eqref{eq:local_stability_LMI_obs_1} implies that, if $\hat v,v\in\mathcal V(\Lambda_\mathrm o)$, then 
    \[\Delta V_\mathrm o\coloneq V_\mathrm o(k+1)-V_\mathrm o(k)<0,\]
    where $V_\mathrm o(k)=\norm{e(k)}^2_{P_\mathrm o}$.
    \item[2.] Next, show that, if \eqref{eq:local_stability_LMI_obs_2}-\eqref{eq:local_stability_LMI_obs_3} are also satisfied, then $e(k)\in\mathcal E(P_{\mathrm{o}}/\gamma)$ implies $\hat v,v\in\mathcal V(\Lambda_\mathrm o)$ and $e(k+1)\in\mathcal E(P_{\mathrm{o}}/\gamma)$ for all $\gamma\in\mathbb{R}_+$ when $ \Lambda_{\mathrm{o}}=0$, and for all $\gamma\in(0,\gamma_{\mathrm{o}}]$ when $ \Lambda_{\mathrm{o}}\neq 0$ and \eqref{eq:obs_regional_constr_1}-\eqref{eq:obs_regional_constr_2} hold.
\end{itemize}
%Define $v=\tilde{A}_\mathrm e\eta+\tilde Bu+\tilde B_ss$, $\hat v =\tilde{A}_\mathrm e\hat{\eta}+\tilde Bu+\tilde B_s\hat s+\tilde LC_\mathrm ee$, and  $\Delta \hat s=\Delta s(\hat v,v)$. 
1. In view of Lemma~\ref{lem:sector_condition} and \eqref{eq:sector_inequality}, for any $\Lambda_\mathrm o\in\mathbb{D}_{\geq0}^\nu$ such that $\Lambda_\mathrm o\prec I_\nu$, if $\hat v,v\in\mathcal{V}(\Lambda_\mathrm o)$, then, for all $S_\mathrm o\in\mathbb{D}_+^\nu$,
\begin{equation*}%\label{eq:sector_inequality_obs}
    (\Delta v-\Delta \hat s)^\top S_\mathrm o(\Delta \hat s-\Lambda_\mathrm o\Delta v)\geq 0,
\end{equation*}
where $\Delta \hat s\coloneq\Delta s(\hat v,v)$ and $\Delta v\coloneq v-\hat v=\tilde A_{\mathrm e,\mathrm L}e+\tilde B_s\Delta\hat s$.
%Therefore, we can write 
%\begin{align*}
%\Delta v&=v-v^-\\
%&=\tilde A_\mathrm e\eta+\tilde Bu+\tilde B_ss-\tilde A_\mathrm e\hat \eta-\tilde Bu-\tilde B_s\hat s-\tilde LC_\mathrm ee\\
%&=\tilde A_{\mathrm e,\mathrm L}e+\tilde B_s\Delta\hat s\,.
%\end{align*}
This condition can be rewritten as
\begin{equation*}%\label{eq:sector_inequality_obs_extended_0}
    (\tilde A_{\mathrm e,\mathrm L}e{+}(\tilde B_s{-}I_\nu)\Delta\hat s)^\top S_\mathrm o(-\Lambda_\mathrm o\tilde A_{\mathrm e,\mathrm L}e+(I_\nu{-}\Lambda_\mathrm o\tilde B_s)\Delta\hat s)\geq 0.
\end{equation*}
Defining $\phi_{\mathrm o}=[e^\top,\ \Delta\hat s^\top]^\top$ and following similar steps to \eqref{eq:sector_inequality_ctr_extended_0}-\eqref{eq:sector_inequality_ctr_extended}, we derive that this condition %\eqref{eq:sector_inequality_obs_extended_0} 
implies
\begin{equation}\label{eq:sector_inequality_obs_extended}
\phi_\mathrm o^\top\left(\begin{bmatrix}0\!\!&\!\!\tilde{A}_{\mathrm e,\mathrm L}^\top S_\mathrm o\\
    S_\mathrm o\tilde{A}_{\mathrm e,\mathrm L}&-U_{\mathrm S,\mathrm o}\end{bmatrix}\!-\!M_{\mathrm o}(\Lambda_\mathrm o)\right)\phi_\mathrm o\!\geq \!0,
\end{equation}
where $U_{\mathrm S,\mathrm o}=(I_\nu-\tilde B_{s})^\top S_\mathrm o+S_\mathrm o(I_\nu-\tilde B_{s})$.\\
Recalling that $A_{\mathrm e,\mathrm L}=A_\mathrm e-LC_\mathrm e$, the observation error dynamics is 
\begin{equation}\label{eq:observation_error_dynamics}
    e(k+1)=A_{\mathrm e,\mathrm L}e(k)+B_s\Delta \hat s(k)\,.
\end{equation}
%Defining $V_\mathrm o(k)=\norm{e(k)}_{P_\mathrm o}^2$ and 
Using \eqref{eq:observation_error_dynamics}, we can write \begin{equation}\label{eq:deltaV_o}
    \begin{aligned}
        \Delta V_\mathrm o%&\!\!=V_\mathrm o(k+1)-V_\mathrm o(k)\\
        &\!\!=e(k+1)^\top P_\mathrm oe(k+1)-e(k)^\top P_\mathrm oe(k)\\
        &\!\!=\!\phi_\mathrm o^\top\left(\begin{bmatrix}
            A_{\mathrm e,\mathrm L}^\top\\B_{s}^\top
        \end{bmatrix}P_\mathrm o\begin{bmatrix}
            A_{\mathrm e,\mathrm L}&B_{s}
        \end{bmatrix}\!\!-\!\!\begin{bmatrix}
            P_\mathrm o &0\\0&0
        \end{bmatrix}\right)\phi_\mathrm o.
    \end{aligned}
    \end{equation}
We can exploit \eqref{eq:sector_inequality_obs_extended} to guarantee $\Delta V_\mathrm o<0$, and therefore that the origin of \eqref{eq:observation_error_dynamics} is asymptotically stable
by imposing
\begin{multline}\label{eq:condition_local_stability_obs}
    \Delta V_\mathrm o+\phi_\mathrm o^\top\left(\begin{bmatrix}0&\tilde{A}_{\mathrm e,\mathrm L}^\top S_\mathrm o\\
    S_\mathrm o\tilde{A}_{\mathrm e,\mathrm L}&-U_{\mathrm S,\mathrm o}\end{bmatrix}{-}M_{\mathrm o}(\Lambda_\mathrm o)\right)\phi_\mathrm o<0,\\\ \forall \hat v,v\in\mathcal{V}(\Lambda_\mathrm o),
\end{multline}
Using \eqref{eq:deltaV_o}, we can rewrite \eqref{eq:condition_local_stability_obs} as  
\begin{multline*}
\phi_\mathrm o^\top\Bigg(\begin{bmatrix}P_\mathrm o&-\tilde{A}_{\mathrm e,\mathrm L}^\top S_\mathrm o\\
    -S_\mathrm o\tilde{A}_{\mathrm e,\mathrm L}&U_{\mathrm S,\mathrm o}\end{bmatrix}-\begin{bmatrix}
            A_{\mathrm e,\mathrm L}^\top\\B_{s}^\top
        \end{bmatrix}P_\mathrm o\begin{bmatrix}
            A_{\mathrm e,\mathrm L}&B_{s}
        \end{bmatrix}\\+M_{\mathrm o}(\Lambda_\mathrm o)\Bigg)\phi_\mathrm o>0, \forall \hat v,v\in\mathcal{V}(\Lambda_\mathrm o),
\end{multline*}
which is satisfied if \eqref{eq:local_stability_LMI_obs_1} holds.\smallskip\\
%We now show that, if \eqref{eq:local_stability_LMI_obs} holds, the condition $e(k)\in\mathcal E(P_{\mathrm{o}}/\gamma)$ implies $e(k+1)\in\mathcal E(P_{\mathrm{o}}/\gamma)$ for all $\gamma\in\mathbb{R}_+$ when $\Lambda_{\mathrm{o}}=0$, and for all $\gamma\in(0,\gamma_{\mathrm{o}}]$ when $\Lambda_{\mathrm{o}}\neq 0$ and \eqref{eq:obs_regional_constr} holds.\\
2. In the trivial case $\Lambda_{\mathrm{o}}= 0$, since $\lambda_{\mathrm o,i}=0$ for all $i=1,\dots,\nu$ by definition, then $\mathcal V(\Lambda_\mathrm o)$ is unbounded and $\hat v,v\in\mathcal V(\Lambda_\mathrm o)$ for all $\hat v,v\in\R^\nu$. Therefore \eqref{eq:local_stability_LMI_obs_1} guarantees that for any $\gamma\in(0,+\infty]$, $e(k)\in\mathcal E(P_\mathrm o/\gamma)$ implies
\(V_\mathrm o(k+1)<V_\mathrm o(k)\leq\gamma,\)
i.e., $e(k+1)\in\mathcal E(P_\mathrm o/\gamma)$.\\  
In the case $\Lambda_{\mathrm{o}}\neq 0$, the same implication holds for all $\gamma\in(0,\gamma_\mathrm o]$, if $e(k)\in\mathcal E(P_\mathrm o/\gamma_\mathrm o)$ implies that $\hat v,v\in\mathcal V(\Lambda_\mathrm o)$. Noting that $\hat v=\tilde A \hat x+\tilde Bu+ \tilde B_s\hat s+\tilde LC_\mathrm ee$, condition \eqref{eq:obs_regional_constr_1} implies that $\hat v\in\mathcal V(\Lambda_\mathrm o)$ for all $e\in\mathcal E(P_\mathrm o/\gamma_\mathrm o)$. Indeed, the set $\mathcal V(\Lambda_\mathrm o)\ominus\tilde LC_\mathrm e\mathcal E(P_\mathrm o/\gamma_\mathrm o)$ is not empty if,  for all $e\in\mathcal E(P_\mathrm o/\gamma_\mathrm o)$ and $i\in\mathcal I(\Lambda_\mathrm o)$, it holds that $|\tilde L_iC_\mathrm ee|\leq \bar v_{i}(\lambda_{\mathrm o,i})$. This condition is satisfied by \eqref{eq:local_stability_LMI_obs_2} in view of~\cite[Lemma~10]{ravasio2025recurrentneuralnetworkbasedrobust}.\\
Similarly, noting that $v=\tilde A \hat x+\tilde Bu+ \tilde B_s s+\tilde A_\mathrm ee$ condition~\eqref{eq:obs_regional_constr_2}  guarantees that $v\in\mathcal V(\Lambda_\mathrm o)$ for all $e\in\mathcal E(P_\mathrm o/\gamma_\mathrm o)$. The set $\mathcal V(\Lambda_\mathrm o)\ominus\tilde A_\mathrm e\mathcal E(P_\mathrm o/\gamma_\mathrm o)$ is not empty if,  for all $e\in\mathcal E(P_\mathrm o/\gamma_\mathrm o)$ and $i\in\mathcal I(\Lambda_\mathrm o)$, $|\tilde A_{\mathrm e,i}e|\leq \bar v_{i}(\lambda_{\mathrm o,i})$. This condition is satisfied by~\eqref{eq:local_stability_LMI_obs_3} in view of~\cite[Lemma~10]{ravasio2025recurrentneuralnetworkbasedrobust}.
\hfill{}$\square$\smallskip\\

% To print the credit authorship contribution details
%\printcredits

%% Loading bibliography style file
%\bibliographystyle{model1-num-names}
\bibliographystyle{cas-model2-names}

% Loading bibliography database
\bibliography{cas-refs}

% Biography
%\bio{}
% Here goes the biography details.
%\endbio

%\bio{pic1}
% Here goes the biography details.
%\endbio

\end{document}